%% file: main.tex
\documentclass[sigconf]{acmart}
\usepackage{popets}

\setcopyright{popets}
\copyrightyear{YYYY}

\acmYear{YYYY}
\acmVolume{YYYY}
\acmNumber{X}
\acmDOI{XXXXXXX.XXXXXXX}
\acmISBN{}
\acmConference{Proceedings on Privacy Enhancing Technologies}
\usepackage{enumitem}
\usepackage{url}
\usepackage{xurl}
\PassOptionsToPackage{hyphens}{url}
\usepackage{hyperref}

\usepackage{breakurl}
\usepackage{tablefootnote}

\newcommand{\parhead}[1]{\medskip\Parhead{#1}}
\newcommand{\Parhead}[1]{\noindent\textbf{#1}\hskip 0.5em\relax}
\usepackage{xspace}
\usepackage{multirow}
\usepackage{comment}
\usepackage{spverbatim}
\usepackage{booktabs}
\usepackage[table]{xcolor}
\usepackage{tablefootnote}
\usepackage{xcolor} 
\usepackage{subfig}
\newcommand{\sys}{CensorLess\xspace}

\begin{document}

\title[\sys]{\sys: Cost-Efficient Censorship Circumvention \\Through Serverless Cloud Functions}

\author{Dayeon Kang}
\orcid{0009-0007-6838-9090}
\affiliation{%
  \institution{University of Massachusetts Amherst}
  \city{Amherst} 
  \state{Massachusetts} 
  \country{USA} 
}
\email{dayeonkang@umass.edu}

\author{Jade Sheffey}
\affiliation{%
  \institution{University of Massachusetts Amherst}
  \city{Amherst}
  \state{Massachusetts}
  \country{USA}}
\email{jsheffey@cs.umass.edu}

\author{Mingshi Wu}
\affiliation{%
  \institution{\href{https://gfw.report}{GFW Report}}
  \city{}
  \country{\rule{0pt}{0pt}}
}
\email{gfw.report@protonmail.com}

\author{Pubali Datta}
\affiliation{%
 \institution{University of Massachusetts Amherst}
 \city{Amherst}
 \state{Massachusetts}
 \country{USA}}
\email{pdatta@umass.edu}

\author{Amir Houmansadr}
\affiliation{%
  \institution{University of Massachusetts Amherst}
  \city{Amherst}
  \state{Massachusetts}
  \country{USA}}
\email{amir@cs.umass.edu}




\renewcommand{\shortauthors}{Kang et al.}

\begin{abstract}
With the increase in Internet censorship globally, various circumvention tools have been designed and developed. However, the monetary cost of these tools deeply impacts both user choice and the sustainability of provider operations. Recent developments in censorship circumvention research attempted to achieve cost efficiency by utilizing Infrastructure-as-a-Service (IaaS) spot instances as bridges, but still incurred substantial expenses related to network connectivity and instance maintenance.

In this work, we present \sys, a circumvention proxy built leveraging the unique benefits of a serverless platform. 
\sys comprises three components:
a local proxy that handles client-side communication and ensures compliance with serverless functions' security restrictions, 
a function refresher that periodically regenerates bridges, 
and a live migration mechanism that maintains continuous connectivity.
\sys inherits the serverless platform’s cost efficiency, ephemerality, scalability, concurrency, and performance.
Compared to existing low-cost, state-of-the-art circumvention techniques, \sys reduces costs by 97\%, while simultaneously enabling robust censorship resistance by employing bridge rotation. 
\end{abstract}


\maketitle

\input{sections/1_introduction}
\input{sections/2_background}
\input{sections/3_system-design}
\input{sections/4_evaluation}
\input{sections/5_discussion}
\input{sections/6_relatedworks}

\input{sections/7_futurework}
\bibliographystyle{ACM-Reference-Format}
\bibliography{main, censor}

\appendix

\input{sections/appendix}









\end{document}

%% file: sections/1_introduction.tex
\section{Introduction}
Internet censorship remains a significant barrier to global information access. To break this barrier, censorship circumvention tools have emerged as a solution. 
The monetary cost of censorship circumvention systems has been a critical factor both for users and system providers. For circumvention tool providers, securing funding is a common and big hurdle because maintaining a circumvention system includes the cost of renting servers, buying upstream bandwidth, continuously rotating IP addresses, and many other tasks~\cite[\S5.3]{Xue2024b}. Open Technology Fund recently increased funding for circumvention tools due to a significant increase in the number of users of these systems~\cite{techeratiLanternUSbacked,dwLeadersMore,opentechSurgeSustain}. The operational cost for large-scale circumvention tools can be prohibitively expensive~\cite{businessplantemplatesWhatOperating}. For example, the Tor project spent \$11,152.72 on a third-party server in February 2018~\cite{torprojecttorprojectSummary}.
Recent surveys with users of circumvention tools ~\cite[\S4.3]{dutkowska2022and},~\cite[\S5.2]{ramesh2023all} show that price is almost always ranked in the top three most important requirements when users select a tool. Especially for users with limited-to-moderate technical expertise, price is a major factor in their decision-making; 71.1\% (436 of 613) of users rank price in their top three~\cite{ramesh2023all}.

Recently, proxy-based circumvention systems have started leveraging cloud computing to reduce operational costs. In proxy-based circumvention tools, a user in the censored region sends traffic to a proxy server that is allowed to access the Internet, and the proxy server routes the traffic to the true destination. The proxy server acts as a bridge or relay for the entire communication, enabling the user to access the Internet freely. The proxy can be positioned across various networks, including Internet Service Providers (ISPs)\cite{Frolov2019b}, Content Distribution Networks (CDNs)~\cite{Zolfaghari2016a}, edge networks~\cite{Nasr2020a, Kon2024a}, and cloud infrastructures~\cite{Brubaker2014a, Kon2024b}. Modern cloud-based circumvention proxies create distributed bridge networks with cloud instances that operate on usage-based pricing models to curb the costs ~\cite{Kon2024b}. 

In this paper, we leverage serverless cloud platforms (also known as Function-as-a-Service or FaaS) to reduce the operational costs of circumvention proxies to a fraction of that of previous approaches. The serverless cloud supports request-based pay-as-you-go pricing, auto-scaling, high concurrency, and ephemerality. The per-request pricing model in serverless platforms reduces cost drastically in designing a circumvention proxy. More importantly, stateless short-lived serverless cloud functions maximize circumvention capability. When the serverless function receives invocation requests, it creates secure, and isolated execution instances (e.g., container)~\cite{amazonWhatLambda, amazonUnderstandingLambda}. These instances are ephemeral (e.g., maximum lifetime of 15 minutes in AWS lambda ~\cite{amazonConfigureLambda}) , and new instances are initiated with new IP addresses. This dynamic, randomized IP-address rotation~\cite{amazonGenerateStatic} complicates IP-based blocking. Additionally, the auto-scaling and high concurrency inherently supported by serverless cloud enable \sys to scale to numerous clients effortlessly. We design \sys, the first serverless censorship-circumvention system that exploits these properties to deliver robust, and extremely cost-efficient circumvention.

Designing a serverless circumvention proxy is challenging because the serverless cloud stack is largely opaque. We are limited in accessing information at layers below the application layer in serverless platforms, causing an inability to conduct socket communication or IP-layer datagram capturing. Disabled socket communication is particularly disruptive because it hinders the capability of the serverless function to act as a proxy. The alternative approach to network communication in serverless cloud requires the use of additional paid cloud services (e.g., API Gateway, Virtual Networks). However, incorporating additional cloud services is at odds with our goal of providing circumvention capability at minimal cost. 

We design \sys to overcome the aforementioned challenges through three core mechanisms:
\textit{request translation}, \textit{automatic bridge generation}, and \textit{live migration}. Since a cheap circumvention tool is our core objective, \sys uses a local proxy server at the client side to avoid socket communication traffic. This proxy server translates Internet traffic into HTTPS requests, allowing users to browse websites transparently despite the limitations posed by serverless functions.
To avoid DNS and TLS SNI-based blocking, \sys regularly retires serverless function bridges and creates new ones across multiple regions. 
To migrate connectivity between bridges and ensure uninterrupted service, \sys seamlessly transfers active network connections, communicating migration information through the current bridge to avoid exposing the new serverless bridge.
By rotating bridges and capitalizing on the ephemeral nature of serverless functions through these three mechanisms, \sys enhances resilience against blocking or filtering by censoring agents. 

Additionally, we make the observation that some serverless environments (e.g., AWS Lambda) provide the capability of domain-fronting ~\cite{Fifield2015a} letting HTTPS requests present an allowed hostname while actually reaching a different backend. We hypothesize this capability persists to allow customers to pair serverless functions with CDN services such as Amazon CloudFront for operational convenience \cite{amazonUsingAmazon}. Because serverless invocations already originate from ephemeral, widely distributed IPs and can be reached via CDN edge domains, domain fronting lets clients present a legitimate-looking TLS SNI while the serverless bridge receives the real request — effectively camouflaging circumvention traffic among benign cloud flows. This design increases the collateral cost of censorship significantly, improves resilience by decoupling the visible and backend hostnames, and avoids additional long‑lived infrastructure. \sys optionally leverages domain fronting in serverless deployments where the local proxy chooses the TLS SNI that appears legitimate and forwards traffic to the serverless bridge, which performs the backend routing.

\sys also supports a more stringent privacy-preserving mode that adopts a Virtual Private Server (VPS) to offer an encrypted communication channel and compatibility with SOCKS proxy, adding minimal cost to vanilla \sys. The private mode has a 20\% cost increase over the vanilla system when it supports 50 serverless proxies. 
In Section~\ref{sec:eval}, our evaluation demonstrates \sys's unparalleled cost-efficiency and performance. Compared to state-of-the-art circumvention proxies, \sys achieves at least 97\% cost savings (63.3\% cost savings for the private mode) and maintains costs below \$3.50 per day, even when scaling to 300 proxies. 
Based on the findings from the experiments in Section \ref{sec:eval}, we provide operational guidelines for maintaining minimal costs when deploying serverless bridges. The contributions of \sys are summarized as follows:
\begin{itemize} [leftmargin=4mm]
    \item We develop \sys , a censorship circumvention proxy that maximizes cost savings without additional overhead, utilizing serverless computing design primitives. 
    \item We implemented \sys on top of AWS Lambda, one of the most popular serverless cloud services, enabling support for other serverless clouds with minimal modifications.
    \item We thoroughly evaluate \sys's performance, censorship circumvention effectiveness, and the costs incurred using standard benchmarks and compare it with state-of-the-art censorship circumvention approaches. \sys reduces costs by at most 97\% compared to the prior least-cost approach, SpotProxy~\cite{Kon2024b}.
\end{itemize}

%% file: sections/2_background.tex
\section{Background and Related Work}

\subsection{Internet Censorship}

Censors around the world have employed 
a variety of techniques and devices 
to inspect and filter network traffic.

\parhead{Website censorship.}
Censors around the world often block websites and Internet services
with a combination of various censorship techniques,
which include,
but not limited to,
IP address blocking~\cite[\S4]{Chai2019a},
DNS injection~\cite{Duan2012a, Hoang2021a, Anonymous2020a, Anonymous2014a}, 
HTTP Host-based filtering~\cite{Rambert2021a}, 
TLS (E)SNI-based filtering~\cite{Chai2019a, Hoang2024a, Bock2020ESNI},
QUIC traffic filtering~\cite{Elmenhorst2021a},
and traffic throttling~\cite{Xue2021a, Anderson2013a}.
These website censorship techniques are operated by diverse governments, China~\cite{Hoang2024a, Zhang2024a, Feng2023a, Bock2021c}, Iran~\cite{Lange2025a, Bock2020a}, Russia~\cite{Kristoff2024a,Ortwein2023a,Ramesh2023a,Xue2022b}, India~\cite{Katira2023a}, Turkmenistan~\cite{Nourin2023a}, etc.
Beyond national enforcement, 
censorship has expanded to regional levels, 
with local governments implementing their own filtering systems~\cite{Zhang2024a, Fan2025a}.

\parhead{Proxy detection.}
More resourceful and capable censors,
including China~\cite{Alice2020a, Wu2023a}, Iran, and Russia,
also identify and block proxy protocols and endpoints,
creating a cat-and-mouse game between censors and netizens~\cite{cat-and-mouse}. 
The censors have documented to target
fully encrypted proxies~\cite{Alice2020a,Wu2023a} (e.g., Shadowsocks~\cite{shadowsocks}, VMess~\cite{vmess}, and Outline~\cite{outline}), 
and TLS-based proxies~\cite{2022-tls-blocking} (e.g., Trojan~\cite{trojan} and Tor~\cite{Winter-obfs2-probe,Winter2012a,knock-knock-tor,Ensafi2015b,Dunna2018a}). 
While China,
Iran,
and Russia have all been employing passive traffic analysis-based censorship techniques to identify proxy protocol~\cite{Wu2023a,Alice2020a,Frolov2019a}, 
China is the only known country that also develops active probing infrastructures and techniques to more accurately identify proxy severs~\cite{Alice2020a,Anonymous2021ShadowsocksAdvise,Anonymous2021ShadowsocksTutorial,Frolov2020b}.

\subsection{Censorship Circumvention}

Censorship circumvention remains a critical challenge for Internet users in regions with restricted access. Over the years, various approaches have emerged to address this challenge, evolving alongside advancements in network technologies and censorship techniques. This section examines the landscape of censorship circumvention tools with different approaches.

\parhead{Proxy-based approaches.}
Proxy-based circumvention tools have been widely and commonly used to counter censorship techniques. These systems place proxy servers~\cite{Douglas2016a, Houmansadr2011a} in uncensored regions, allowing clients to access the Internet through these intermediaries.
Although relatively accessible to non-technical users, conventional censorship circumvention tools such as VPNs~\cite{ferguson1998vpn, zhang2004overview} often feature static IP addresses that are easily identified and blocked by censors.
Researchers have experimented with placing proxies in various network environments, such as Internet Service Providers (ISPs)~\cite{Frolov2019b}, Content Distribution Networks (CDNs)~\cite{Zolfaghari2016a}, and edge networks~\cite{Nasr2020a, Kon2024a} to improve resilience against detection. 
More sophisticated approaches leverage peer-to-peer architectures~\cite{fiat2002censorship}, such as MassBrowser~\cite{Nasr2020a}, which uses normal users (i.e., volunteers) as proxies, or Tor~\cite{dingledine2004tor}, which implements multi-hop routing to provide stronger anonymity through circuit-based routing over multiple nodes. 

\parhead{Cloud-based circumvention systems.}
Prior works have leveraged unique characteristics of cloud services to design proxies in circumvention tools. 
CloudTransport~\cite{Brubaker2014a} pioneered the use of cloud infrastructure to resist censorship by tunneling network traffic through encrypted cloud storage services. This approach utilizes the feature that public cloud storage systems provide a popular encrypted medium accessible from both inside and outside censor-controlled networks, making it difficult for censors to distinguish circumvention traffic from legitimate storage access. 

While CloudTransport focused on using cloud products as encryption tunnels, SpotProxy~\cite{Kon2024b} directly employed IaaS Virtual Machines (VMs) as bridges.
Kon et al. used spot instances as proxies and implemented a function rejuvenation component that repeatedly searches for cheaper instances, creates new VMs or changes IP addresses, and migrates proxies accordingly. 
To maintain continuous connections during these transitions, 
they adopted Network Address Translation (NAT) techniques that temporarily relay traffic while migration takes place. 
Although SpotProxy effectively reduces costs through using spot instances compared to always-on instances, 
it introduces considerable operational overheads 
attributed to repeatedly searching for cheaper instances, 
implementing two types of rejuvenation (IP and instance), 
maintaining NAT infrastructure, 
and enduring longer configuration times for new instances. 
\sys addresses these inefficiencies by leveraging serverless computing to reduce operational complexity.

\parhead{HTTP proxy.}
HTTP proxies~\cite{weaver2014here} represent one of the most widely deployed proxy types in circumvention systems due to their simplicity and compatibility with standard web traffic. These proxies operate at the application layer, relaying HTTP requests from clients to destination web servers and returning responses. Traditional HTTP proxies operate on dedicated servers with static IP addresses, making them vulnerable to discovery and blocking by censors. Modern circumvention systems have enhanced HTTP proxying~\cite{pandiaraja2015web, perino2018proxytorrent} by implementing features such as request encryption and traffic obfuscation to improve resilience against detection. HTTP or SOCKS proxies offer a particularly accessible approach, allowing users to access blocked content without installing specialized software by submitting URLs into the
browser’s settings window. Despite continuous countermeasures by censors, HTTP proxying remains fundamental to many circumvention systems due to its flexibility and relative simplicity of implementation.

\sys builds upon this proxy mechanism while addressing its limitations by adopting core features of the serverless cloud architecture, providing a more cost-effective and resilient approach to HTTP-based circumvention.

\parhead{Domain fronting.}
Domain fronting~\cite{Fifield2015a} represents a sophisticated evolution of HTTP proxying that uses Content Delivery Networks (CDNs) to circumvent Internet censorship. This technique allows HTTPS requests to appear as if they are destined for permitted domains while actually accessing blocked content.
The core mechanism of domain fronting relies on using different domain names across TLS and HTTP layers. Normally, the HTTPS \texttt{Host} header, DNS query, and TLS Server Name Indication (SNI) extension have the same domain name. In domain fronting, a connection uses identical domain names in both the DNS query and the SNI of the TLS layer while employing a different domain in the HTTP \texttt{Host} header. This strategic inconsistency prevents censors from identifying the actual destination, as they can only monitor the DNS request and TLS SNI extension but cannot inspect the hostname in the encrypted HTTP layer. Upon receiving the request, the frontend server uses the hostname from the HTTP \texttt{Host} header to properly route communications to the covert destination.

\section{Why use Serverless Computing?}

Serverless computing or Function-as-a-Service (FaaS)~\cite{mcgrath2017serverless} represents a modern cloud computing paradigm characterized by event-driven request execution with infrastructure management abstracted from developers. Serverless cloud shifts resource management, load balancing, function deployment, and execution responsibilities from developers to cloud service providers~\cite{van2018serverless}. This approach enables developers to deploy stateless functions (i.e., we call serverless functions, which are function-level scripts) that automatically scale based on request load. Unlike the traditional Infrastructure-as-a-Service (IaaS) model (e.g., Amazon EC2, Google Compute Engine, and Azure Virtual Machines), the user does not have to manage the underlying hardware and software stack, and works with a high-level abstraction. Major cloud providers, including Amazon AWS, Google Cloud, and Microsoft Azure, have embraced serverless architecture, offering pay-as-you-go pricing models that typically charge around \$0.20 per million invocations, with variations based on memory allocation, storage requirements, and execution duration. The combination of resources managed by providers, fine-grained scaling, cost efficiency, and streamlined development has positioned serverless computing as an increasingly important paradigm in cloud computing. While previous projects have utilized serverless functions as proxies—ranging from simple implementations to research-oriented reverse proxies for machine learning applications~\cite{mahmoudi2022mlproxy}—the potential of serverless in building effective circumvention tools remains unexplored. The inherent characteristics of serverless cloud present both opportunities and challenges in designing censorship circumvention proxies. In this section, we discuss the opportunities and challenges of using serverless cloud functions in developing \sys an inexpensive censorship circumvention proxy.   


\subsection{Merits of Serverless Computing} \label{sec:motivation-merits}

The stateless nature of serverless functions means all resources —including IP addresses— are ephemeral by design. Serverless functions are typically accessed via URLs or REST-based API calls rather than direct IP addresses, and their underlying IP addresses rotate unpredictably~\cite{amazonGenerateStatic}. 
This characteristic fundamentally complicates censors' efforts to track or block users based on IP addresses and reduces the risk of exposing sensitive information about clients in censored regions.

Another notable property of serverless functions is that they are event-driven. 
Unlike traditional IaaS clouds that charge based on the entire period of renting and running compute instances regardless of usage, serverless platforms incur costs only when functions are invoked and executed. 
This event-driven model eliminates the need to maintain continuously running instances or to invest effort in finding lower-cost instances. For users in censored regions and circumvention tool providers, for whom cost is a critical factor, this request-based pricing model presents a notable advantage over traditional proxy approaches.

Serverless platforms offer autoscaling that automatically adjusts to function request volumes, deploying event-triggered function instances to handle concurrent requests. This capability enables flexible handling of usage spikes without manual intervention and reduces operational overhead for circumvention tool providers while ensuring consistent service availability during periods of high demand.

Cloud providers typically offer serverless functions across numerous global regions, enabling rapid deployment of proxies across diverse geographic locations. This distribution capability enhances resilience against regional blocking attempts and allows for strategic positioning of proxies to optimize performance for users in specific censored regions.

\subsection{Design Challenges} \label{sec:motivation-challenge} 

While the ephemeral nature of serverless functions provides notable censorship resistance advantages, it also imposes significant development constraints.

\begin{itemize}[leftmargin=4mm]
\item The serverless software-stack is largely opaque to the tenants restricting the developer's access within the function instance (e.g., containers). Consequently, network visibility is restricted at the application layer, preventing the use of socket connections, packet capturing, kernel-level operations, etc. While cloud providers offer additional services like API Gateways, static IP addresses, and virtual networks to address these limitations, adopting these services diminishes the cost advantages of serverless computing. Thus, the primary challenge in building \sys is implementing effective network communication using only application-layer libraries while maintaining cost efficiency, ultimately leading us to utilize HTTP GET and POST methods as detailed in Section \ref{sec:client-side}. 
\item When a serverless function hasn't been recently invoked, the instances hosting that function are removed from the deployment environment, and the cloud platform may need to initialize its execution environment before processing new requests, introducing "cold start" latency~\cite{amazonUnderstandingLambda}. This latency can impact user experience, particularly for time-sensitive browsing activities.
\item Serverless platforms impose constraints on payload sizes that may impact proxy functionality. AWS Lambda limits payloads to 6 MB (extendable to 20 MB under certain conditions)~\cite{amazonLambdaQuotas}, Google Cloud Functions caps uncompressed HTTP sizes at 10~MB (32~MB in the second generation)~\cite{googleQuotasCloud}, and Microsoft Azure Functions, while technically unlimited in response size, has practical limits around 250~MB~\cite{microsoftAzureFunctions}. These constraints necessitate careful design considerations for handling larger web content. We implemented \sys in an AWS environment and used Node.js to extend the payload size limit to 20~MB.
\end{itemize}


We carefully address the aforementioned challenges through innovative design patterns in building \sys. To the best of our knowledge,  \sys is the first censorship circumvention solution that uses the inherent ephemerality in serverless cloud to achieve cost-efficiency and resilience to censorship.

%% file: sections/3_system-design.tex
\section{System Design}

\begin{figure*}[h]
  \centering
  \includegraphics[width=0.8\linewidth]{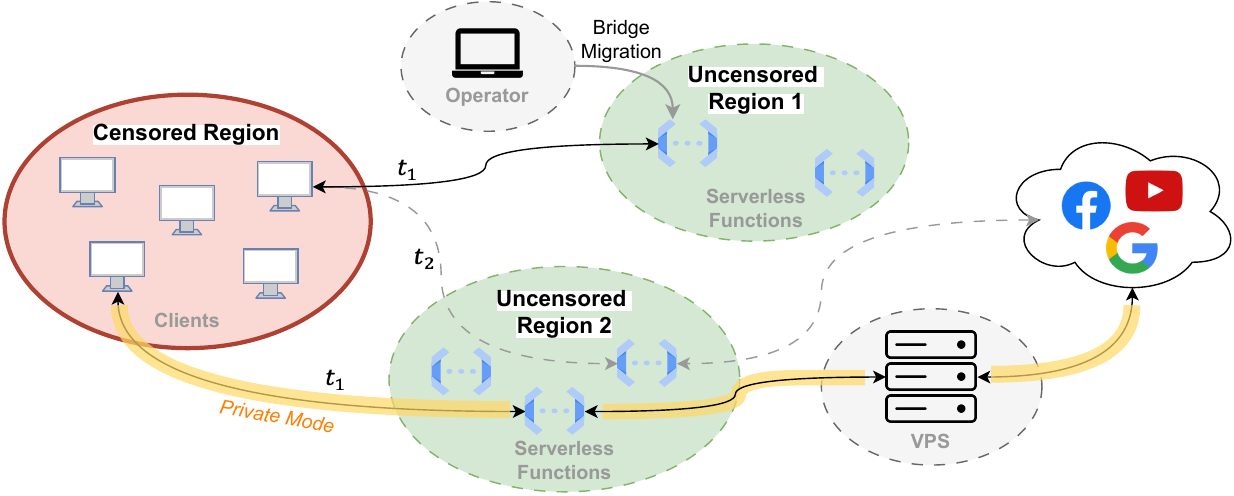}
  \caption{Overview of \sys operational workflow}
  \label{main-fig}
\end{figure*}
In this section, we describe the design of \sys, a serverless circumvention proxy.

\subsection{Design Goals}
We designed \sys with the following goals:
\begin{itemize}[leftmargin=4mm] 
    \item \parhead{Cost reduction:} \sys is designed to minimize costs by providing a proxy service using only serverless functions, without requiring additional paid services offered by cloud providers, such as API Gateway or Virtual Networks.
    \item \parhead{Censorship resistance:} To succeed as a censorship circumvention tool, \sys is designed to effectively resist traffic tracking by IP addresses and blocking via DNS and TLS SNI through bridge rotation. 
    \item \parhead{Scalability:} \sys delivers robust concurrency support, enabling service expansion from individual users to large-scale organizations while maintaining consistent performance.
\end{itemize}

\subsection{Threat Model}~\label{sec:threat-model}
We assume that \sys's clients are located in censored regions, and the serverless function, which acts as a proxy, is placed in a region that has not been censored. As a trust assumption for other components in \sys, operators must not disclose client-bridge mappings or inspect/record traffic in vanilla mode, and VPSs used for private mode must forward traffic correctly without logging. The censorship authority observes Internet traffic from the client and can block or interfere with the traffic if the client attempts to access censored websites. 

The censor possesses the capability to inspect, manipulate, and block network traffic flowing within its jurisdiction. We consider every agent in our system to be rational. The censor is external to the cloud provider and acts as a rational adversary who aims to minimize collateral damage when implementing blocking measures. For instance, the censor is unlikely to block entire cloud provider IP ranges or domain names due to the significant collateral damage such actions would cause to legitimate services.
Censors can also act as clients of \sys in an attempt to identify bridges and disrupt the service. They may actively probe suspected bridge endpoints and attempt to enumerate individual serverless function instances. 

We also assume that the cloud provider acts rationally~\cite{Kon2024b, Nasr2020a}, meaning they are willing to serve users as long as doing so does not expose them to any risk. For example, the cloud service provider may disclose encrypted network communications under coarse-grained security policies. In the real world, there is a possibility that cloud providers may behave irrationally, such as colluding with censors due to legal or economic pressure. We have discussed this scenario in Section~\ref{sec:sec-discussion}.

We do not assume the censor has the ability to compromise end-user devices (e.g., by installing monitoring software), as this would undermine any privacy-enhancing technology. While censors might employ deep packet inspection (DPI) to identify traffic patterns associated with circumvention tools or block entire cloud providers such as Russia, the ephemeral nature of serverless bridges presents significant challenges to such identification efforts.


\begin{figure}[h]
  \centering
  \includegraphics[width=\linewidth]{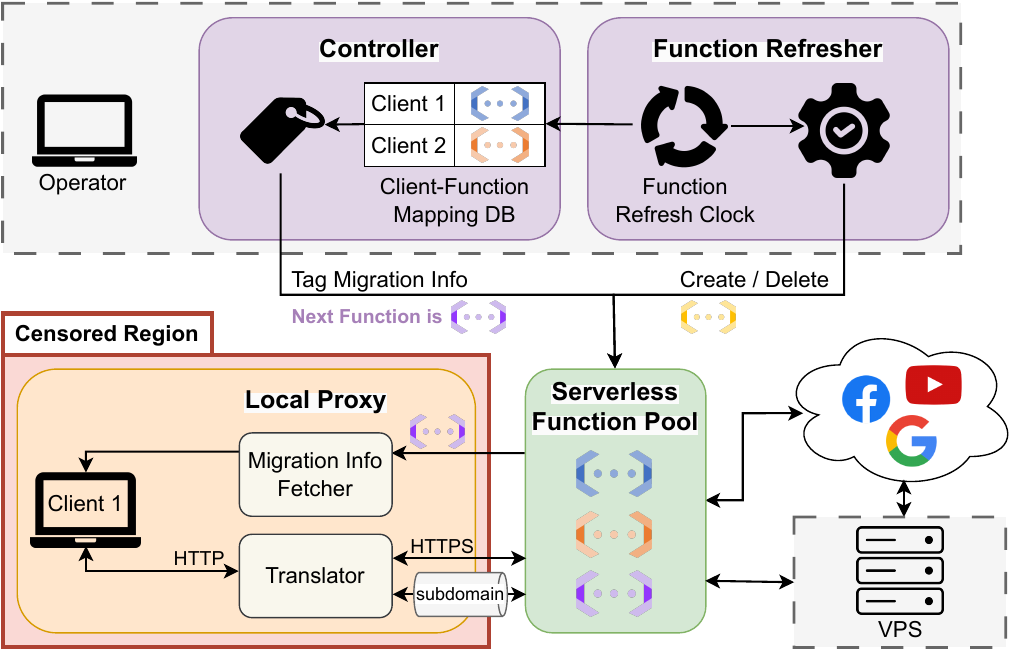}
  \caption{Overview of \sys architecture}
  \label{system-design}
\end{figure}

\subsection{Overview} \label{sec:overview}
\sys circumvents censorship by establishing ephemeral serverless computing-based bridges between clients in censored regions and their intended destinations. As Figure~\ref{main-fig} illustrates, clients in a censored region send requests to serverless functions located in an uncensored region at time $t_1$. 
Serverless function bridges in uncensored regions inspect the request and act as proxies, forwarding the traffic to the client's intended destination. The destination treats the serverless bridges as the actual clients and sends data back to them, then the function returns it to the original clients.
The operator--any individual or organization willing to pay per-invocation costs--can be located in any region with cloud service access, managing serverless function bridges and orchestrating connectivity migration.
When time reaches $t_2$ after a certain period, clients change their bridge to different serverless functions in random uncensored regions to continually obfuscate traffic patterns from censors. 

\sys provides an additional component for privacy. The private mode operates with reduced cost efficiency while preserving the privacy of the client's requests. Based on the user's preference and their priority, the user can select between vanilla \sys and private mode. When the user selects the private mode, the client sends encrypted messages using a SOCKS proxy to a serverless function bridge over HTTPS. Then, the serverless bridge relays the message to the Virtual Private Server (VPS), and the VPS forwards the encrypted message to the destination on the internet. Except for sending encrypted messages and additionally connecting the VPS as a bridge to the Internet, the rest of the workflow is the same as the vanilla \sys, such as the operator's serverless bridge management and bridge rotation.

Taking a closer look at this system, it consists of three key mechanisms that addresses the key serverless-based design challenges, and four modules that support these key mechanisms. Key mechanisms are request translation, automatic bridge generation, and live connectivity migration. We will address these mechanisms in detail by grouping them into the client-side and the operator-side processes in Sections \ref{sec:client-side} and \ref{sec:operator-side}. The end-to-end workflow of the private mode is described in Section~\ref{sec:private-mode}.

Our system comprises four primary modules as represented in Figure~\ref{system-design}: controller, function refresher, serverless function pool, and local proxy. The \textit{controller} and \textit{function refresher} are managed by the operator and work together to orchestrate network connectivity migration between serverless bridges. 
Operating in coordination with the controller, the function refresher manages the lifecycle of serverless bridges. It periodically generates new serverless functions in batches across diverse regions, applies appropriate security configurations, and notifies the controller of their availability. It can adjust the scalability of the serverless function pool by defining the size of serverless bridges. 
The controller maintains a Client-Function mapping database that tracks serverless bridge assignments for each client.
When the controller updates this database, it informs clients about their new bridge assignments by tagging information needed to migrate onto the current serverless function.

The \textit{serverless function pool} contains batches of serverless bridges. Each bridge modifies incoming request headers from clients and composes new outgoing packets to appear as though the serverless bridge itself is the client. The private mode parses the HTTPS payload and forwards it as a full request. When the serverless bridge receives responses from the client's destination, it returns them to the client without modification. These serverless functions are lightweight (147~MB Docker image for vanilla system) as they serve only as minimal bridges. 

The \textit{local proxy} handles most request modifications to reduce processing overhead at the bridge.
The biggest obstacle to the local proxy is the fundamental restrictions of FaaS. Since FaaS does not allow functions to communicate with a socket, the local proxy addresses these restrictions by establishing the application-level connections with the serverless bridge and adopting protocol conversion. 
The local proxy comprises two components: (1) a translator that converts 
client requests to HTTPS for serverless compatibility, using SOCKS proxy 
in private mode, and (2) a migration information fetcher that periodically 
detects and seamlessly transitions to new bridge URLs.

\subsection{Client Side} \label{sec:client-side}
A client starts \sys by installing the local proxy server, which is preconfigured with an initial serverless bridge URL. The serverless bridge is periodically updated in the background. After the client runs the local proxy, network traffic is relayed to the allocated serverless function bridge, and the intended data from the target destination is returned through the bridge. For transparent HTTPS communication with target websites, the local proxy server adopts a key mechanism, request translation. 

\parhead{Request translation.}
Request translation addresses the fundamental limitation of serverless functions: their inability to support socket-based communications, which are commonly used today. To overcome this constraint, \sys converts HTTP requests in the local proxy server using two principal components: a \textit{translator} and a \textit{migration information fetcher}.

The translator intercepts client HTTP GET and POST requests while listening on a designated port to avoid the HTTPS CONNECT method, which uses socket communication, and processes them for secure transmission to the serverless bridge. Given the security vulnerabilities inherent in standard HTTP requests, the translator converts them to HTTPS, modifying or removing irrelevant request headers in the process. 
The converted traffic is then forwarded directly to the assigned serverless function bridge through an HTTPS-encrypted tunnel. At the serverless bridge, minimal request modifications are performed to emulate an actual endpoint, reinitiate the HTTPS tunnel, and send traffic to the intended destination. Due to the inherent restrictions of serverless functions and the need to minimize costs without extra cloud services, the communication protocol is separated.
When responses are received from the destination via the serverless bridge, the translator performs the reverse conversion, transforming HTTPS responses back into HTTP format for client consumption. This process includes removing unnecessary HTTPS-related headers and modifying response bodies when required.
In private mode, a SOCKS proxy is used, and the encrypted message captured from the client is sent to the serverless bridge over an HTTPS tunnel by the translator.
Additionally, the local proxy implements streaming responses and pipelining to enhance network performance.

To enhance resistance against detection by censors, the client exclusively communicates with its allocated serverless bridge rather than directly interacting with other system components. The migration information fetcher periodically checks for migration tags embedded in responses from the current bridge. When a new bridge URL is detected, it seamlessly migrates the connectivity without requiring external components or introducing communication overhead. This HTTP-based approach enables transparent and immediate migration for clients while maintaining continuous service.

\subsection{Operator Side} \label{sec:operator-side}
To effectively obfuscate traffic patterns and enhance resistance against detection, \sys regularly rotates serverless bridges across various uncensored regions. The operator executes two key mechanisms to facilitate this rotation: automatic bridge generation and live migration.
The system starts with a client registration to \sys through a local proxy server installation by a client.
Upon registration, the operator randomly selects the next serverless bridge from the available pool, allocates it to the client, and records this assignment in the Client-Function mapping database. This initiates the ongoing cycle of bridge rotation and management.

\parhead{Automatic bridge generation.}
The function refresher operates according to operator-defined time periods, creating new batches of serverless functions and removing those that have passed the operational cycle.
When generating new serverless bridges, it employs a lightweight Docker image and applies strict security policies that limit access permissions to the minimum required for operation. Once new bridges are deployed, the function refresher notifies the controller via HTTP request, enabling subsequent client migration. After clients have migrated away from bridges at the end of their cycle, the function refresher automatically removes these deprecated functions, ensuring that no bridge remains active long enough to be identified and targeted by censorship agents. The period until the bridge is identified in practice is expected to be 2 days, as reported by Fifield et al.~\cite{Fifield2016a}; Tor bridges in China are discovered within 2-36 days.

\parhead{Live migration.}
The controller orchestrates the migration process by managing the Client-Function mapping database. Upon receiving notification of new bridge availability from the function refresher, the controller updates the mapping database, removing old bridge assignments and allocating new ones to clients. This database update triggers the migration process. 
Under the stateless nature of serverless functions, the controller implements a tag-based migration mechanism. Rather than directly communicating migration instructions with clients, the controller attaches the URL of newly assigned bridges as tags. These tags are then detected by the client's migration information fetcher, enabling seamless transition to the new bridge.
The old bridge is automatically removed by the function refresher after clients have completed their migration. Throughout this process, the controller maintains minimal communication with clients, primarily interacting with a serverless bridge. 
This design simplifies the migration process and reduces detectable communication patterns. The HTTP-based implementation enables this efficient live migration without requiring complex protocols or procedures.
Through these coordinated client-side and operator-side mechanisms, \sys achieves continuous service while regularly altering network pathways, effectively circumventing SNI-based censorship. 


\subsection{\sys for Privacy}~\label{sec:private-mode}
\sys provides public internet connectivity, circumventing censor agents, at minimal cost. However, it poses a risk that an irrational cloud provider could potentially access data at a serverless bridge side. Therefore, we present a private mode that preserves privacy for \sys. Adopting the private mode introduces a trade-off between cost/performance and privacy/anonymity. We compare the additional cost incurred by this trade-off in Section~\ref{sec:evaluation-cost}. While the privacy for clients against malicious operators at the serverless bridge is preserved, malicious operators at the VPS may analyze ingress and egress traffic patterns. We address this limitation in Section~\ref{sec:sec-discussion}.

The privacy-preserving module deploys a Virtual Private Server (VPS) for the SOCKS protocol, long-lived connections, and message encryption, which increases operational costs. As shown in Figure~\ref{main-fig}, VPS is located after the serverless bridge and before the target destination in the uncensored region. Like the vanilla \sys, the connection between the client and the serverless bridge uses HTTPS, but the communication between the serverless bridge and the VPS server uses TCP. 

\parhead{Encryption model.}
To protect the data in the message via encryption, the client and the VPS server each have a public key and private key ($C_{pr}, C_{pub}, S_{pr}, S_{pub}$), and they share their public keys in the bootstrap step, coordinated by the operator. When a client sends a message to a server that contains an encrypted payload ($E(m)$) with the destination, the client encrypts the message with the server's public key ($c_{req} \gets E(S_{pub}, E(m))$), and sends it to the server through the serverless bridge as an HTTPS payload. The server decrypts the message with its private key ($E(m) \gets D(S_{pr}, c_{req})$) and validates the incrementing nonce included in the message. When the server sends the request to the target destination, the connection ID ($ConnID$) is granted by the server, and it is signed with the server's private key ($DS \gets Sign(S_{pr}, ConnID)$). After the server receives the encrypted response ($E(r)$) from the target destination, the server attaches the signed connection ID to the encrypted response and encrypts the entire message with the client's public key ($c_{res} \gets E(C_{pub}, DS|E(r))$). Finally, the client receives the response by decrypting the server's message using its private key ($E(r) \gets D(C_{pr}, c_{res})$).

With this encryption model, four security properties are maintained. Since we use the Ed25519~\cite{Ed25519} public-key signature system for encryption, the client and server are authenticated with Ed25519 public keys configured out-of-band. To prevent replay attacks, we adopt a monotonically increasing nonce. The signed connection ID proves that the client owns the connection, and all encrypted payloads secure confidentiality. A formal proof of the security properties of private mode is left for future work.  

\parhead{End-to-end communication.}
The client initiating the communication is a SOCKS proxy, and it forwards the message to the serverless bridge, embedding it into an HTTPS payload as a vanilla \sys to comply with the restrictions of the serverless function. The request (i.e., HTTPS payload sent by the client) includes the VPS server address in a SOCKS address format to make the connection with the server and the payload that is encrypted with the server's public key. When the serverless bridge receives this HTTPS request from the client, the bridge establishes a TCP connection with the VPS server, and it parses and forwards the HTTPS payload as a message. When the server receives the TCP segment from the bridge, it decrypts the payload using its private key and verifies the incrementing nonce for the public key it was given. Next, it either initiates TCP connections or ferries data between the client via the serverless bridge's connections or the target destination. The server buffers responses for each client that is identified by the public key until the client requests them through another serverless bridge invocation. Buffered responses are controlled and cleaned up with a timeout and a maximum buffer size.

When the server receives the response from the target destination, it encrypts the response using the client's public key and sends it back to the serverless bridge over the TCP connection. The serverless bridge places the TCP segment into an HTTPS payload and sends the request over HTTPS. Then, the client parses the payload and decrypts the actual response from the target destination with its private key. Finally, the client can route the data to the SOCKS communication channel. 

\subsection{Domain Fronting on \sys} \label{sec:domain-fronting}
To enhance its censorship circumvention capabilities, \sys can optionally leverage domain fronting supported by certain serverless platforms. It is  
a technique that conceals the true destination of HTTPS traffic by utilizing the discrepancy between different layers of communication protocols~\cite{Fifield2015a}.
While major CDNs including AWS, Google, and Microsoft have withdrawn their support for domain fronting practices, largely to preserve their commercial relationships with censoring states~\cite{torprojectDomainFronting, thevergeAmazonServices, signalLetterFrom, amazonEnhancedDomain, microsoftSecuringApproach}, 
we discovered that domain fronting continues to be available for AWS Lambda. We tested the domain fronting on Microsoft Azure functions, but it does not allow a mismatched SNI and Host. By providing the additional option for connection establishment through domain fronting in \sys, we improve the censorship resistance of \sys by exploiting the fact that the censor has to pay a high cost to block serverless functions.

\parhead{Mechanism.}
\sys adapts the traditional domain fronting for the serverless environment. 
In this approach, a client connects to a permitted cloud domain (e.g., \texttt{*.lambda-url.*.on.aws} for AWS Lambda) 
and sends an encrypted HTTPS request with a different \texttt{Host} header representing the censored destination.






This process involves
(1) locally establishing the connection to the targeted destination domain, 
(2) setting the HTTP \texttt{Host} header within the encrypted channel to the specific serverless function bridge, 
and (3) utilizing the cloud provider's content delivery infrastructure to route the request appropriately. 
The local proxy server selects TLS SNI values to ensure that 
the domain fronting requests appear legitimate to both censorship systems and cloud provider infrastructure.
This subdomain may be the same as in other serverless functions in the same region, even a completely fabricated subdomain unregistered with AWS.

This implementation is particularly effective because censors face a significant dilemma--blocking all traffic to major cloud providers would cause substantial collateral damage to legitimate services and applications,
and the ephemeral serverless endpoints further complicate filtering efforts.

\parhead{Enhanced resilience through serverless.}
Unlike traditional domain fronting implementations that rely on a limited set of front domains~\cite{Fifield2015a}, \sys can distribute requests across numerous serverless functions deployed across different regions and namespaces within the cloud provider's domain space. This distribution significantly increases the cost of comprehensive blocking.

Even if AWS Lambda decides to disable domain fronting, \sys, without domain fronting, still maintains resilience against identification and filtering attempts by dynamically selecting the domains of serverless functions across multiple regions. Additionally, \sys allows operators to customize the domain selection interval, strengthening the system's robustness against censorship efforts. Incorporating domain fronting alongside \sys' censorship resistance mechanisms provides even greater blocking resistance by preventing actual front-end exposure and increasing the cost to block the system.

%% file: sections/4_evaluation.tex
\section{Implementation}
We implemented \sys to validate our design and demonstrate its practicality. Our implementation consists of approximately 2,000 lines of code written in TypeScript, Node.js, and Python. While we primarily targeted AWS as our deployment platform due to its widespread adoption, the architecture is designed to be cloud-agnostic and can be deployed on alternative cloud providers with minimal modifications to the serverless function management components.

Our implementation incorporates several technical optimizations to ensure minimal latency and overhead. We implemented HTTP response streaming with request pipelining to reduce latency and improve throughput, particularly for web browsing applications. By distributing processing responsibilities between the local proxy and serverless bridge, we minimize the computational overhead at the serverless function layer, where execution time affects costs.

\parhead{Serverless function bridge.}
The serverless function bridge represents the core of our circumvention system. We implemented this component with particular attention to minimizing resource consumption and execution time. The bridge was developed using Node.js with a codebase of less than 100 lines and containerized as a 146.7~MB Docker image to facilitate rapid deployment. We limited dependencies to only four external libraries, \texttt{axios}, \texttt{https}, \texttt{stream}, and \texttt{aws-sdk}, while implementing minimal request processing logic to maintain efficiency. The bridge also integrates a tagging mechanism for migration coordination by separately handling tag requests and general HTTP traffic within a single function.

To ensure security, each serverless function bridge operates according to the principle of least privilege. The security policy assigns only the essential AWS permissions required for operation: Amazon~Resource~Group~Tagging~API:TagResources and GetTagValues~\footnote{\url{https://docs.aws.amazon.com/service-authorization/latest/reference/list_amazonresourcegrouptaggingapi.html}}, Elastic~Container~Registry:GetRepositoryPolicy~\footnote{\url{https://docs.aws.amazon.com/service-authorization/latest/reference/list_amazonelasticcontainerregistrypublic.html}}, and Lambda:InvokeFunctionUrl, CreateFunction, and ListFunctions~\footnote{\url{https://docs.aws.amazon.com/service-authorization/latest/reference/list_awslambda.html}}. 
This restricted permission set minimizes potential security vulnerabilities while maintaining full functionality.

The function refresher creates new serverless functions using the Docker image, deploys functions across diverse geographical regions, applies security policies to each created function, and removes expired bridges after completion of their usage cycle using the AWS Software Development Kit \texttt{boto3}. Access to serverless bridges and the use of \texttt{boto3} can only be performed by operators who own the serverless function pool.

\parhead{Local proxy server.}
Clients interact with the system by running a local proxy server application implemented in TypeScript (under 400 lines of code). The local proxy server manipulates request headers from a user-defined port to convert HTTP requests to HTTPS, overcoming the restrictions of serverless platforms. It necessarily accepts the following header names: 'Cookie', 'User-Agent', 'Host', 'Content-Type', 'Permission-Policy', 'Accept', 'Accept-Encoding', and 'Accept-Language'. If the 'Strict-Transport-Security' header is included, the local proxy changes the value to 'max-age=0', and if 'Referer' or 'Origin' headers contain 'http:', it replaces it with 'https:'. After accepting the necessary headers, the local proxy redefines the \texttt{Host} as the serverless bridge and adds a new header called 'X-Host', which indicates the true destination. Finally, the local proxy sends an HTTPS request to the serverless bridge with the reassembled headers. 
For adopting domain fronting, it sends requests to a subdomain with the true serverless bridge \texttt{Host} header. 
When the local proxy receives a response from the serverless bridge, it filters the acceptable request headers for consistency for the user and streams the data using pipelining. If the data includes 'https:', the local proxy replaces the string with 'http:'.

\parhead{Protocol for private mode.}
As the private mode establishes a connection with the VPS server and uses a SOCKS proxy, it requires a protocol for an HTTPS request payload. The HTTPS payload is a message that communicates with the VPS server, and it includes the VPS server address in SOCKS address format, the server port, the payload length, and the encrypted payload. The encrypted payload consists of the client's public key, a nonce to prevent replay attacks, and the message. The VPS server interprets the message, which can be of several types. For a ``start connection'' message, the message contains the target address and the target port. A ``data'' message contains a signed connection ID, a compression flag to indicate whether the data is compressed or not, the original data length, the compressed length, and the data. A data message of length 0 is considered a ``keepalive'' message. A ``close connection'' message contains only a signature.

When it comes to the response message from the VPS server to the client encrypted with the client's public key, the response message also follows a defined message structure. For the ``connection establishment'' message, the response carries the type and connection ID. ``Data response'' message has the type, connection ID, compression flag, original data length, actual length, and data, and ``connection close'' message has the type, connection ID, message length, and message. If an error occurs, the message conveys the type, connection ID, error code, message length, and message. 

In addition, the VPS filters private IP ranges to prevent Server-Side Request Forgery (SSRF) attacks. Connection attempts to the addresses in the loopback range, private networks, link-local addresses, multicast, broadcast, documentation, and unspecified ranges are rejected with an error code. 

\section{Evaluation}\label{sec:eval}
This section demonstrates that our implementation of \sys achieves effective censorship circumvention while maintaining reasonable performance and low cost.
We evaluated our serverless computing-based censorship circumvention system across four critical dimensions: live migration performance, serverless function performance as a bridge, operational costs, and censorship circumvention efficiency. 

\subsection{\sys Performance} 

\begin{figure*}[t]
    \centering
    \subfloat[Browsing website (cnn.com) with lots of objects.]{
        \includegraphics[width=0.3\textwidth]{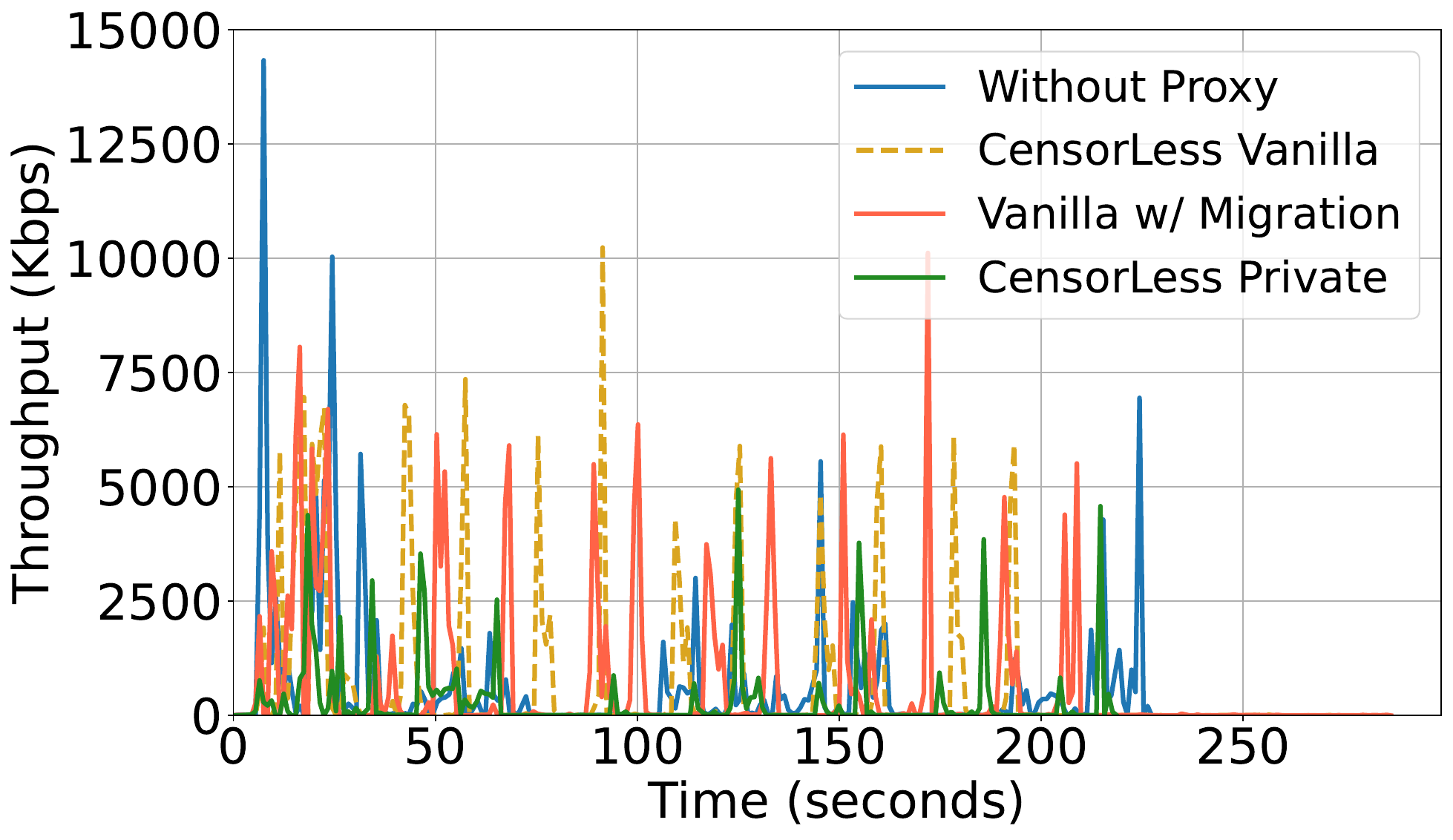}
        \label{fig:throughput_cnn}
    }
    \hfill
    \subfloat[Downloading PDF files.]{
        \includegraphics[width=0.285\textwidth]{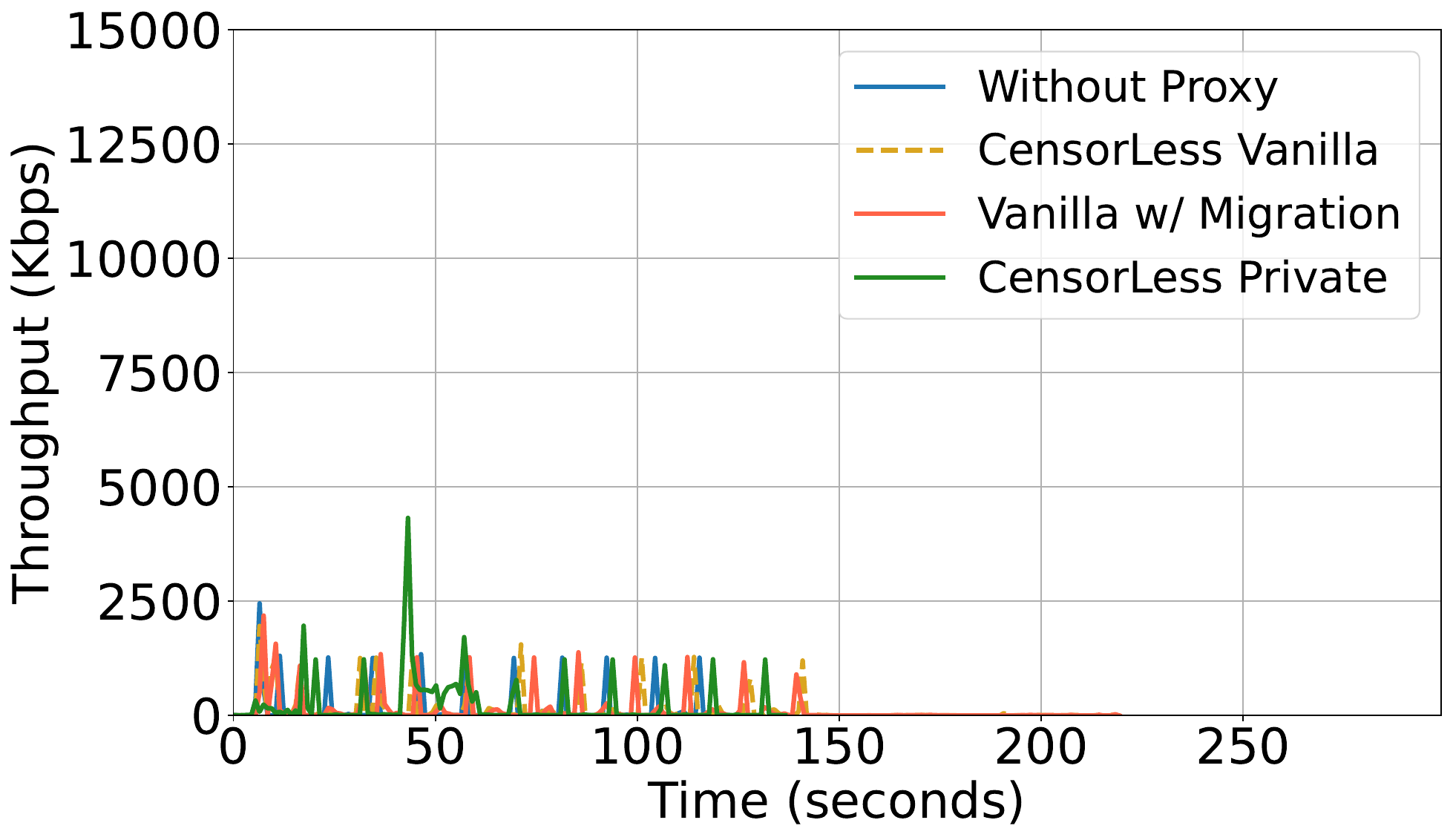}
        \label{fig:throughput-pdf}
    }
    \hfill
    \subfloat[Watching short streamed video.]{
        \includegraphics[width=0.3\textwidth]{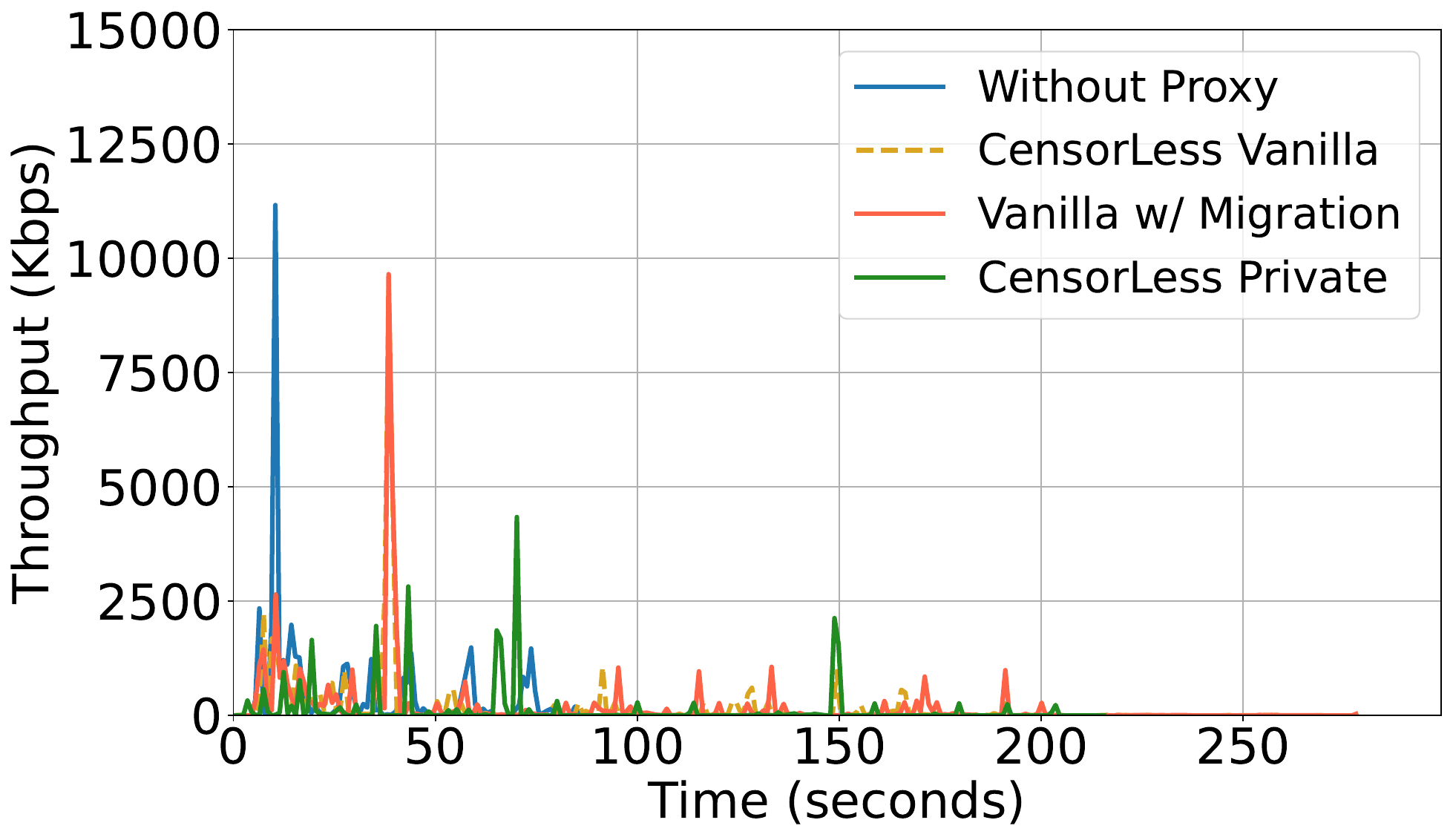}
        \label{fig:throughput-streaming}
    }
    \caption{This diagram shows throughput patterns according to different content loadings. The user browsed a news website and read 5 articles of different lengths (Figure~\ref{fig:throughput_cnn}), browsed a research paper and downloaded the PDF file 10 times (Figure~\ref{fig:throughput-pdf}), and the user watched the same video 5 times (Figure~\ref{fig:throughput-streaming}). } 
    \label{fig:throughput}
\end{figure*}

To evaluate our systems' capability, we measured throughput with a single user under different content-loading scenarios with a Selenium script for consistency: reading articles from a web page with numerous objects, PDF downloads, and streaming video with actual browsing and interaction (clicking, downloading, and watching video). We compared the system's performance between without \sys, static serverless bridges in vanilla mode, bridges migrated every 20 seconds with a single client, and static serverless bridges in private mode. 
In practice, the migration cycle can be set longer than 20 seconds, consistent with the observed blocking period (in Section \ref{sec:circumvention-effect}).
Figure~\ref{fig:throughput} reveals that \sys vanilla mode, regardless of migration, takes a longer time compared to without proxy and private mode in terms of the total execution time, even though the private mode requires more time for receiving the first response due to the secure connection setup process. The load-time waterfall graph for these use cases and each \sys mode is in Appendix~\ref{appendix:eval-performance}. 
Although requests from vanilla mode \sys route via the serverless bridge, it follows closely to the non-proxy throughput pattern, with negligible overhead and without significant performance penalty, while sacrificing privacy. Private mode creates sustained traffic even when the system is not actively loading content to maintain the SOCKS communication channel by polling.  
Comparing migration and non-migration scenarios, there are no considerable throughput pattern differences, with only slight increases in total execution time during migration periods. Since our system uses HTTPS to communicate with the serverless bridge, the migration does not require much time or processing steps to move the connectivity, and with and without migration show similar patterns across all three cases.

\subsection{Serverless Function Bridge Performance}
\begin{figure}[h]
  \centering
  \includegraphics[width=\linewidth]{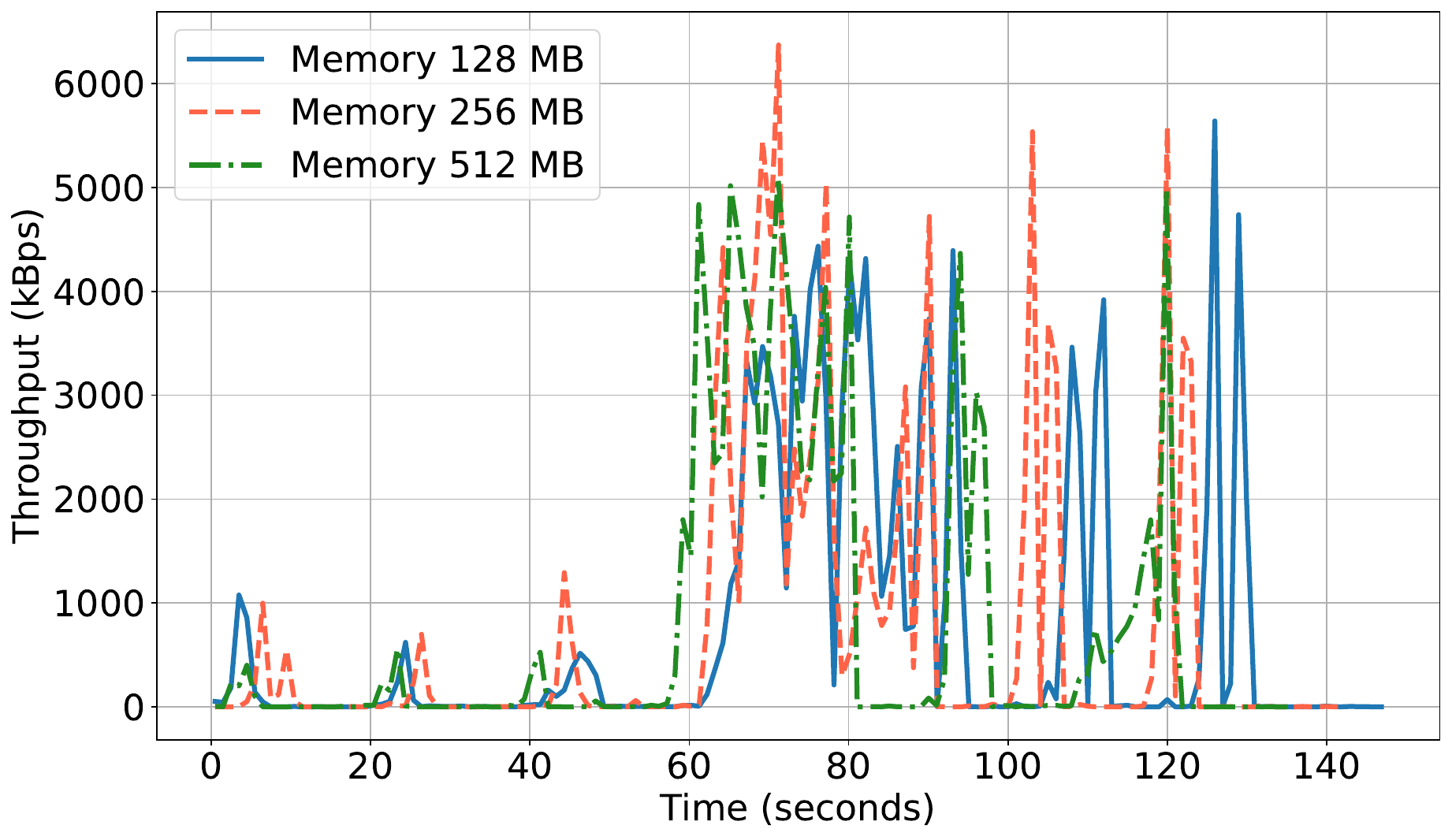}
  \caption{\textbf{Results of throughput by different serverless function bridge configurations.} These configurations have different memory sizes, ranging from 128~MB to 512~MB, with the same timeout (15~seconds) and ephemeral storage size (512~MB).}
  \label{fig:serverless-config}
\end{figure}

Our system's performance directly correlates with serverless function configuration. We evaluated three key configuration parameters: timeout, memory size, and ephemeral storage size.
Setting a 15-second timeout proved optimal for all experiments, as shorter timeouts couldn't properly load streaming content while longer timeouts provided no considerable performance benefits. Similarly, ephemeral storage size affects temporary storage capacity rather than performance when memory allocation remains constant.

However, memory allocation impacts execution time. We measured throughput by repeating the same tasks (browsing a search engine, browsing a news website, and watching a short streamed video three times each) with different serverless function memory configurations. We conducted evaluations with varying memory allocations while keeping all other configurations constant (e.g., 15-second timeout and 512~MB default storage size). As shown in Figure~\ref{fig:serverless-config}, serverless functions with larger memory allocations (512~MB) demonstrated shorter total execution times compared to smaller allocations (128~MB, 256~MB) when performing identical tasks. This occurs because larger memory allocations allow the serverless function to store and process more data simultaneously. However, the serverless function bridge with a 15-second timeout, 512~MB default storage size, and 128~MB default memory size is sufficient in practical use, which does not cause conceivable degradation of performance and usability.


\begin{figure}[h]
  \centering
  \includegraphics[width=\linewidth]{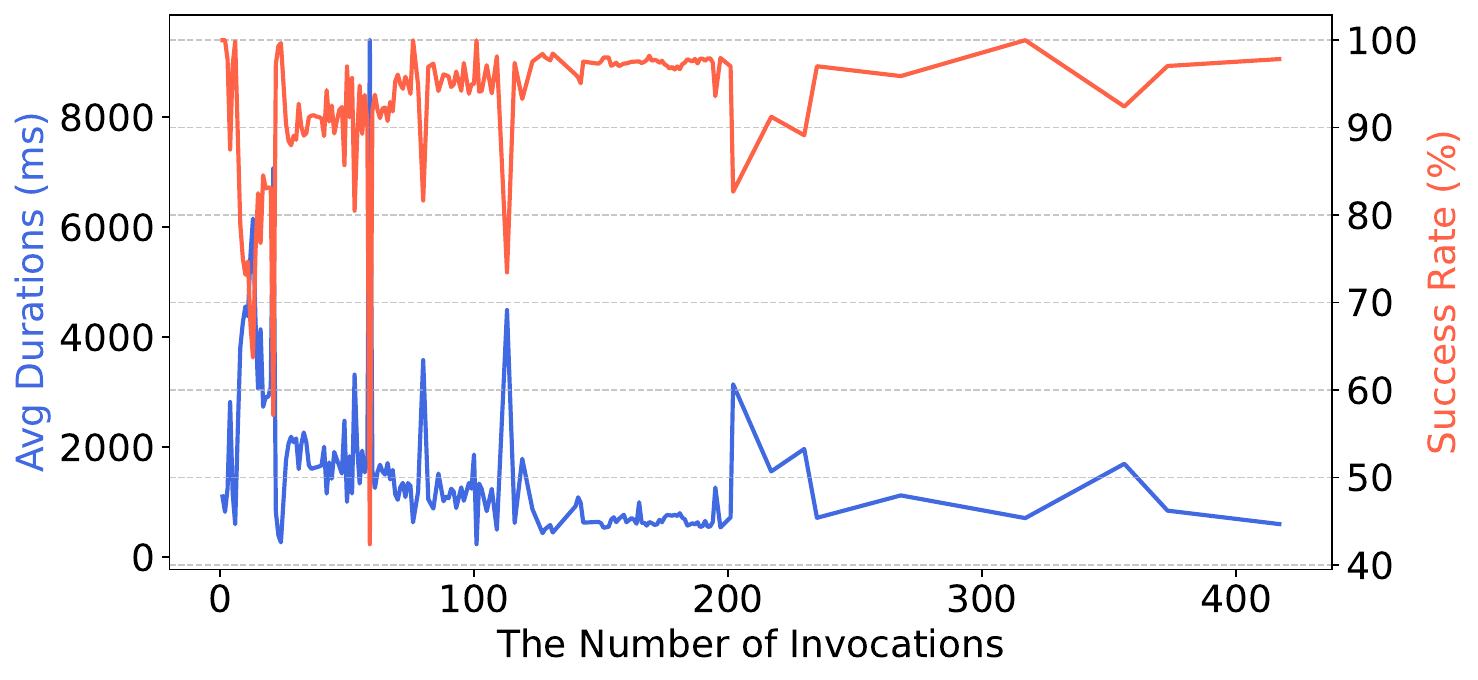}
  \caption{\textbf{Results of the serverless function bridge's concurrency experiment.} Based on our experimental log, we sorted the average duration and success rate according to the specific number of invocations.}
  \label{fig:concurrency}
\end{figure}

One of FaaS's greatest advantages is concurrency through autoscaling. Due to concurrency and autoscaling, serverless functions can support a stable service, and \sys is no exception. We evaluated our system's reliability under concentrated request loads on one serverless bridge by analyzing average duration and success rates as concurrency indicators based on past experiment logs.
As shown in Figure~\ref{fig:concurrency}, during the initial phase (approximately 0-50 invocations), we observe notable volatility in both metrics. The average duration spikes to over 6000~ms in some functions, while the success rate occasionally drops below 90\%. This initial instability can be attributed to the cold start phenomenon common in serverless environments, where the first invocations to a newly migrated bridge require additional time for container initialization and deployment. The system stabilizes considerably between invocations 100-200, with average durations consistently staying below 1000~ms and success rates maintaining above 95\%. This demonstrates the effectiveness of the FaaS platform's autoscaling capability, which dynamically allocates resources to handle the increasing load. Although a very high request load may happen only infrequently, \sys's serverless function bridge implemented on AWS Lambda by default can support a stable service even when concentrated burst requests reach 1000 concurrent invocations \cite{amazonUnderstandingLambda}. 
This inherent scalability makes our system reliable under a large amount of load.

\subsection{Operational Costs} \label{sec:evaluation-cost}
\begin{figure}[h]
  \centering
  \includegraphics[width=\linewidth]{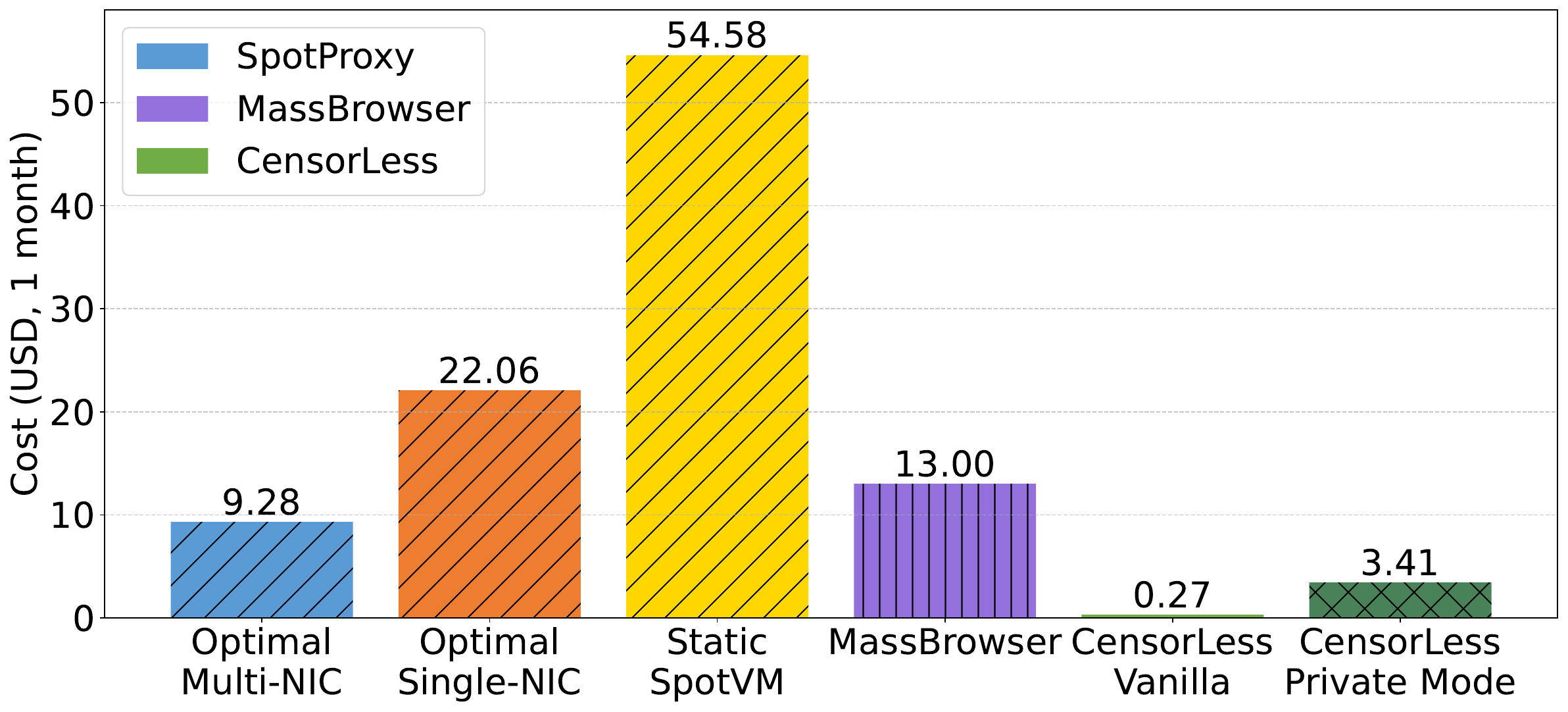}
  \caption{\textbf{Cost comparison (1 month) results of a single proxy.} Optimal multi-NIC, Optimal single-NIC, and Static SpotVM are from the SpotProxy paper\cite{Kon2024b}. As SpotProxy used actual operating costs, we calculated our cost based on our AWS bill and their assumptions.}
  \label{fig:cost-comparison}
\end{figure}

Cost efficiency is our system's highest priority.  
We compared monthly operational costs against existing economical circumvention tools, SpotProxy~\cite{Kon2024b} and MassBrowser~\cite{Nasr2020a}. 
SpotProxy presents three scenarios of using spot instances: using the cheapest spot instances with single-NIC, with multi-NIC (they assumed 3 NICs, which is approximately 3 times cheaper than single-NIC), and using a static spot instance. Unlike MassBrowser, which requires persistent VMs for the operator, both SpotProxy and \sys stem primarily from their reliance on cloud services, SpotProxy on time-based pricing and \sys on request-based pricing. 

For a consistent comparison, we followed SpotProxy's cost analysis assumption~\cite[\S11.1]{Kon2024b}, which is that each proxy processes 6.76 GB of traffic monthly. We then compare monetary costs based on each system's published cost analysis methodology and pricing model~\cite{calculatorPricingCalculator}. 
For our system, we calculated costs based on actual AWS Lambda billing, 0.00296 MB/request used. With 6.76~GB generating approximately 2,338,885~requests (6.76~GB $/$ $($0.00296~MB$/$request) = 2,338,885.14~requests) and AWS Lambda charging \$0.20 per million requests, including 1~million free requests per month, our monthly cost is only \$0.27 (with 1000~ms duration, 128~MB memory, and 512~MB storage). 
Since the private mode of our system additionally uses a VPS (EC2 in AWS), we used the \texttt{t4g.micro} (2~CPUs and 1~GB memory) EC2 instance. Based on the AWS pricing calculator, VPS costs \$3.14 with EC2 Instance Savings Plans. In total, the monthly cost of the private mode is \$3.41 (\$0.27 + \$3.14). 
As Figure~\ref{fig:cost-comparison} demonstrates, our approach achieves remarkable cost efficiency--34.4~times cheaper (97.1\% savings) and even private mode, which is our most expensive mode, is 2.72~times cheaper (63.3\% savings) than SpotProxy's most economical multi-NIC configuration.

\begin{figure}[h]
  \centering
  \includegraphics[width=\linewidth]{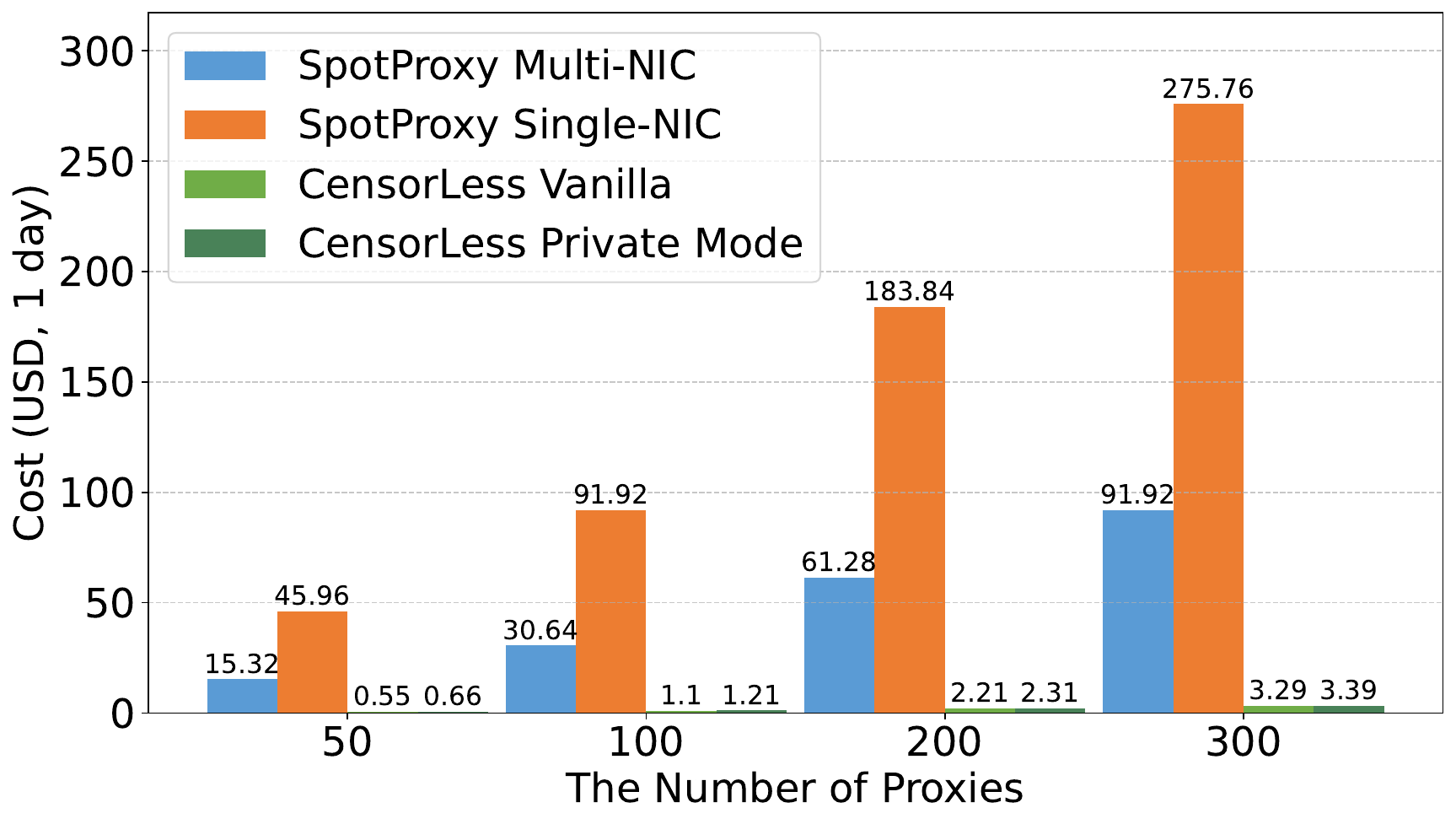}
  \caption{\textbf{Cost comparison (1~day) between SpotProxy and \sys when the number of proxies increases.} Since SpotProxy uses an hour-based pricing model and the vanilla system uses a request-based pricing model, the costs increase at a constant rate. (Private mode additionally charges an hour-based VPS.)}
  \label{fig:cost-scale}
\end{figure}

We evaluated our system's cost scalability against SpotProxy. We assumed that each proxy handles one request per second sequentially for one hour, reflecting a scenario similar to SpotProxy's cost analysis~\cite[\S11.1]{Kon2024b}. The spot instance used for SpotProxy is an \texttt{m6g.large}, the cheapest instance (at an hourly spot VM price of \$0.0383) capable of running the SpotProxy program. SpotProxy is charged based on the hours the instances are in use, while our system is charged based on the number of requests. The number of proxy instances and requests increases at a constant rate. For the private mode, there is an additional charge for the single VPS (\texttt{t4g.micro} hourly costs \$0.004). When the private mode serves with a single proxy, individual requests may be slower. The operator can mitigate this slower speed by adopting additional VPS instances, and the charged amount will increase. Figure~\ref{fig:cost-scale} shows that our systems still cost less than \$3.50 even with 300 proxies. This suggests that leveraging FaaS is a more cost-effective option than using the cheapest IaaS service. Even with private mode, adopting more VPS instances to increase speed or enhance privacy still costs less than an IaaS service, with only a gradual and small linear increase (we demonstrate this in Appendix~\ref{appendix:eval-cost}). 

\subsection{Circumvention Effectiveness}\label{sec:circumvention-effect}
\subsubsection{\textbf{Simulation}}

\begin{figure}[h]
  \centering
  \includegraphics[width=\linewidth]{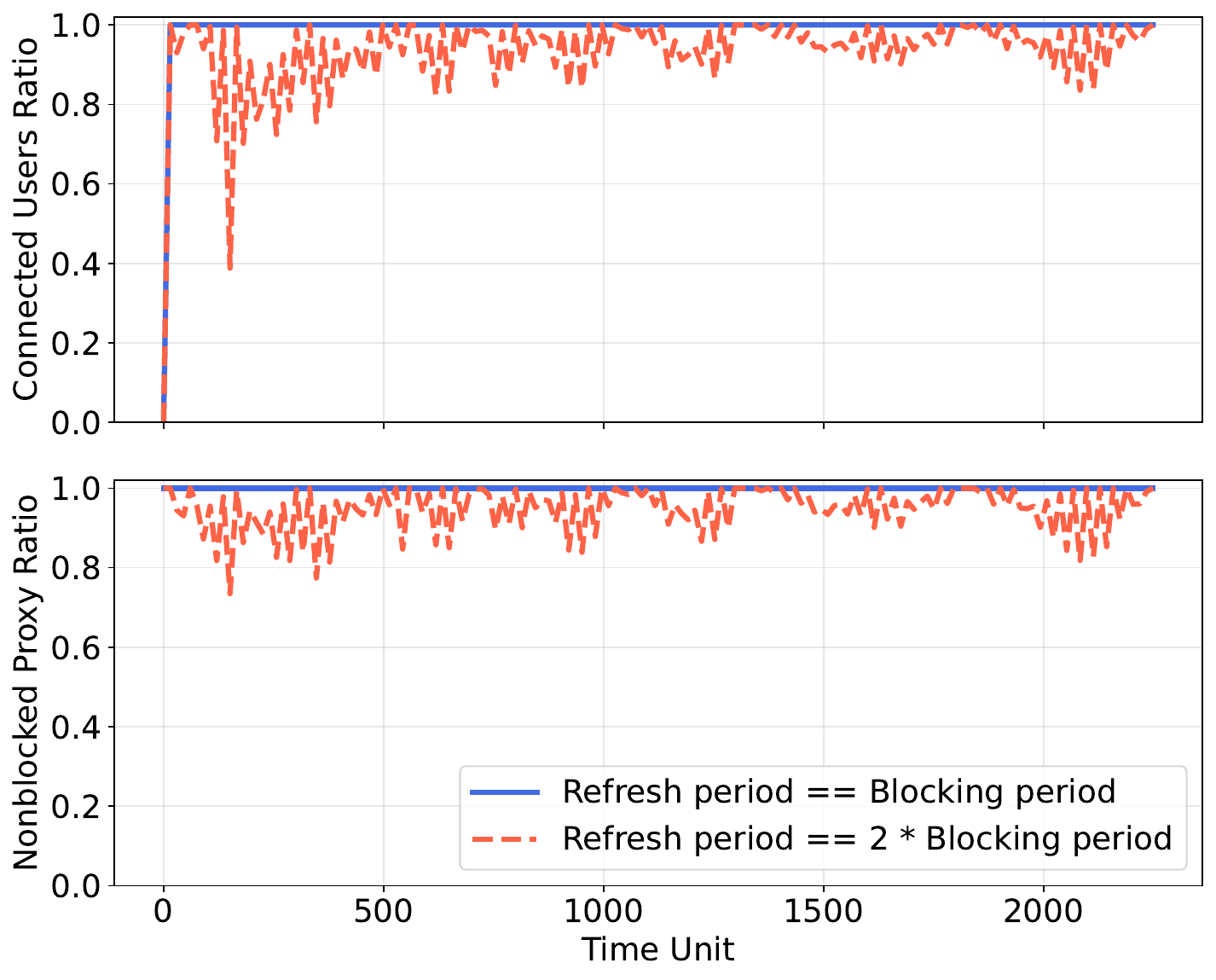}
  \caption{\textbf{Censor simulation using the Optimal method.} \sys preserves around 95\% of connected users and nonblocked proxies, even when 50\% of total clients are censor agents (when the refresh cycle is twice the censorship cycle).}
  \label{fig:simulation-optimal}
\end{figure}

We evaluated our URL refreshing method's censorship effectiveness using the state-of-the-art game-theoretic simulation framework for censorship circumvention proposed by Nasr et al.~\cite{nasr2019enemy} 
This simulator employed three utility functions: client utility function ($U_{a_i}^t(p_x)$), proxy utility function ($U_{p_x}^t(a_i)$), and censor utility function ($U_C^t) $ for client $a_i$, proxy $p_x$, and censor $C$.
The client utility function (\ref{eq:client_utility}) is a weighted sum of the proxy importance factors of the number of users who know the proxy ($B_{p_x}^t$), the number of users connected to the proxy ($c_{p_x}^t$), the total time utilization of the proxy $\tau_{p_x}^t$ and client locations ($d_{a_i, p_x}$). $\beta_1$, $\beta_2$, $\beta_3$, and $\beta_4$ are scaling factors that set the relative importance of proxy metrics, but we set all scaling factors to 1 in our evaluation.
\begin{equation}
U_{a_i}^t(p_x) 
= \beta_1 B_{p_x}^t 
+ \beta_2 c_{p_x}^t 
+ \beta_3 \tau_{p_x}^t 
- \beta_4 d_{a_i,p_x}
\label{eq:client_utility}
\end{equation}

The proxy utility function (\ref{eq:proxy_utility}) is a weighted sum of the client metrics, proxy utilization ($T_{a_i}$), the number of requests for new proxy addresses ($R_{a_i}^t$), the number of assigned blocked proxies ($\gamma_{a_i}^t$), the number of blocked proxies that a user knows ($\delta_{a_i}^t$), and client locations ($d_{a_i, p_x}$) with the objectives that assigning as many, unblocked, and long lived proxies. We set the scaling factors, $\alpha_1$, $\alpha_2$, $\alpha_3$, $\alpha_4$, and $\alpha_5$, to 2, 1, 1, 2, and 10, respectively.
\begin{equation}
    U_{p_x}^t(a_i) = (\alpha_1min(T_{a_i}, \overline{T}) - \alpha_2R_{a_i}^t - \alpha_3\gamma_{a_i}^t - \alpha_4\delta_{a_i}^t - \alpha_5d_{a_i, p_x})\label{eq:proxy_utility}
\end{equation}


The censor utility function (\ref{eq:censor_utility}) balances two competing objectives: proxy discovery and blocking impact. 

\begin{equation}
U_C^t = \omega \sum_{a_i \in J} U_C^t(a_i) + r_{\text{Blocked}}\label{eq:censor_utility}
\end{equation}

$\sum_{a_i \in J} U_C^t(a_i)$ represents the aggregate proxy discovery capability across all censoring agents $J$, $r_{\text{Blocked}}$ denotes the fraction of censored clients unable to obtain working proxies, and $\omega$ is a weighting factor indicating the censor's strategic preference. The utility of censoring agent $a_i$, $U_C^t(a_i)$, is the average of $a_i$'s scores by proxies (i.e., $\mathbb{E}_{p_x \in P}[U_{p_x}^t(a_i)]$). In the implementation, we followed SpotProxy's blocking capability that censors the proxy after 2 time units after discovery.  

We applied these utility functions with minor implementation changes because our clients are assigned serverless bridge URLs, while this simulator distributes proxies to clients based on IP addresses. 
We assumed that each serverless bridge is paired with a single IP address, and the blocked URL cannot be used again, considering SNI-based blocking. Thus, our blocking mechanism is identical to IP-based blocking. To implement URL-based proxy assignment, we followed the serverless function's URL-generation rule; a string of fixed length randomly mixed with letters and numbers for a unique URL, particularly in AWS Lambda. 



We evaluated our system's circumvention effectiveness by the ratio of connected users and non-blocked proxies against the optimal censor. The optimal censor uses an optimal game-theoretic approach as its censorship strategy, such as adjusting its waiting times for discovering new proxies and the frequency or intensity of its blocking efforts to maximize censoring effectiveness, aiming to uncover more proxies and block more clients. Setting 50\% of clients as censor agents, we compared scenarios where refresh periods were either equal to or double the blocking periods.
Every hyperparameter and configuration followed the SpotProxy's optimized values used for their evaluation~\cite[\S8]{Kon2024b}, and we compared the circumvention efficacy in Appendix~\ref{appendix:eval-simulation}.

Against optimal game-theoretic censors (Figure~\ref{fig:simulation-optimal}), our system demonstrates even more stable and higher connectivity across both client and proxy dimensions, even when using less frequent refreshing. To maximize reliability, we recommend setting migration cycles at least as frequently as expected censoring periods. 



\begin{table}[!t]
    \caption{Website access results of \sys in the censored region of Nanjing, China.}
    \label{tab:china-censor}
    \centering
    \renewcommand{\arraystretch}{1.3}
    \begin{tabular}{lcc}
        \toprule
        \textbf{Websites} & \textbf{Vanilla} & \textbf{Private Mode} \\
        \midrule
        google.com.et & \checkmark\tablefootnote{It indicates the system successfully receives an HTTP status code from the website.} & \checkmark \\
        huffpost.com & \checkmark & \checkmark \\
        thecitizen.in & \checkmark & \checkmark \\
        chatgpt.org & \checkmark & \checkmark \\
        gamestorrents.fm & \checkmark & \checkmark \\
        voachinese.com & \checkmark & \checkmark \\
        podcasts-online.org & \checkmark & \checkmark \\
        biblechat.ai & \checkmark & \checkmark \\
        chatdoc.com & \checkmark & \checkmark \\
        vpn.com & \checkmark & \checkmark \\
        \bottomrule
    \end{tabular}
\end{table}

\subsubsection{\textbf{In Censored Region.}}
We experimented with how \sys successfully circumvents censorship in the actual censored region of Nanjing, China. 
To test domain blocking, we collected 500 blocked domains in mainland China from the 5,000-domain DNS A experiment conducted by Sheffey et al.~\cite{sheffey2025ll} The blocked domains tested included websites for search engines, AI tools, social media, news, entertainment, content sharing, pornography, etc.
The current Chinese censorship pattern prevents users from receiving content from these websites over time. We sent 2 requests to each domain using a \texttt{curl} command, and measured whether the website returned an HTTP status code or no response within 100 seconds. We considered a website censored if there was no response within 100 seconds to explicitly distinguish between censored and response delay (usually, the status code was returned within 15 seconds). If the response was returned within 100 seconds, we sent the request to the next domain. 
Table~\ref{tab:china-censor} shows the results for 10 websites, with the 50 results provided in Appendix~\ref{appendix:experiment}, and the total results are available in our artifact.
As shown in Table~\ref{tab:china-censor}, both the \sys vanilla mode and the private mode receive responses to every request sent to blocked domains (the private mode fails to tubegays.xxx; we addressed the reason for failure in Appendix~\ref{appendix:experiment}, which is due to the website policy; reject request from AWS VPS). 

This experiment was conducted in limited settings: a client of \sys was located in Autonomous System Number (ASN) AS45090 of Alibaba Cloud network in Nanjing, China, not in the residential networks for ethical reasons. During 36+ hours of in-region testing, our system, with 50 different serverless functions, was not blocked.

%% file: sections/5_discussion.tex

\section{Security Discussion \& Limitations}~\label{sec:sec-discussion}

In this section, we discuss possible attack scenarios against \sys and corresponding defenses. 

\parhead{HSTS preloading.} 
Serverless is at the core of designing \sys, and one inherent limitation of serverless is that providers prevent socket communication. Consequently, vanilla \sys relies on HTTP requests intercepted locally and translated to HTTPS. This approach encounters complications with modern websites that implement HTTP Strict Transport Security (HSTS)~\cite{dolnak2017introduction}. Web browsers maintain HSTS preloading lists that automatically redirect HTTP requests to HTTPS for specified domains before any network communication occurs. This client-side enforcement prevents \sys from capturing these requests through the local proxy, blocking access to websites on the HSTS preloading list. 

There are two solutions to address the HSTS preloading issues. First, clients use \sys private mode, sacrificing the cost benefit of the vanilla system. Since the private mode local proxy uses a SOCKS proxy and establishes a TCP connection with the VPS, the client does not encounter the browser's force redirect from HTTP to HTTPS.
Second, websites can be accessed through an earlier version of a web browser that does not have an HSTS preloading list.
In early April 2025, when we conducted our experiments, every website was accessible in Firefox without HSTS preloading issues.




\parhead{Cloud provider as an irrational actor.}
In Section~\ref{sec:threat-model}, we assume cloud providers act rationally and will avoid measures that cause large collateral damage to their customers. We acknowledge, however, that a provider could instead collude with a censor and adopt countermeasures such as destination-based blocking or detection via traffic/operator-behavior pattern analysis. Monitoring the target destination (toward the public internet) could not effectively distinguish \sys's from other traffic because serverless functions are commonly used with third-party APIs on the internet or as web crawlers. Blocking via traffic analysis can be prevented with \sys deployed in private mode with multiple serverless bridges. 
While censors could limit \sys by detecting operator-behavior patterns (e.g., repeated function creation), this threat can be mitigated by varying the serverless bridge creation period and adopting the current bridge's URL update.
Effectively, aggressive blocking aimed at \sys would also disrupt other cloud‑based circumvention tools (e.g., Shadowsocks~\cite{shadowsocks},  V2Ray~\cite{v2ray}) and many benign services, creating strong economic and reputational disincentives for providers. For these reasons, we model the provider as a rational actor. Moreover, \sys follows and is implemented based on the documented serverless features of Amazon Web Services (AWS).

\parhead{Protocol support and HTTPS decapsulation.}
Due to restrictions of serverless functions on WebSocket communication, \sys supports only HTTPS GET and POST methods in vanilla mode, despite the widespread use of WebSocket connections in the modern web. Consistent with this design, \sys decapsulates HTTPS requests at the serverless bridge. These limitations are mitigated in the private mode with a tradeoff of slightly increased operational cost.
In the private mode, we employed a custom proxying protocol that persistently tracks connection state across serverless bridge invocations rather than standardized approaches such as HTTP CONNECT and MASQUE, using the fact that serverless only accepts traffic via HTTP and stream responses within reasonable time/data bounds. 

\parhead{Traffic analysis attack.}
Censors can analyze HTTP patterns via modified headers or traffic patterns, but request headers sent to serverless functions are user-defined and variable. Identifying \sys traffic is difficult unless the cloud provider inspects payloads; \sys with private mode prevents such payload exposure through encryption. Since serverless functions are widely used as website domains or web crawlers, CensorLess traffic does not have unique patterns. IP-based blocking is also challenging for censors because serverless functions share public IP addresses. In future work, we will explore fingerprinting resistance techniques, such as User-Agent standardization~\cite{torprojectFingerprintingProtections}.

\parhead{Blocking via domain filtering.}
Censors could implement filtering based on characteristic domain patterns of serverless functions. For instance, AWS Lambda function URLs contain distinctive strings for the specific usage that could be used for blanket blocking. However, the widespread legitimate use of serverless functions for business applications creates substantial collateral damage from such broad filtering, making this approach less likely in practice. 
If such filtering emerges, techniques like DNS over HTTPS (DoH)~\cite{bumanglag2020impact} could provide an additional layer of obfuscation by encrypting DNS queries through an HTTPS encrypted session.

\parhead{DoS attack.}
The attacker may send a large amount of network connection requests to the service to cause financial exhaustion, exploiting the request-based pay-as-you-go pricing model of the serverless function. This type of attack is a Denial-of-Service (DoS) attack that can occur in a serverless environment. Any serverless function is vulnerable to DoS, and the cloud provider throttles the number of invocations as a countermeasure. At the service level, we mitigate DoS attacks by adding a client authentication step before user registration.

\parhead{Cold start attack.} 
Serverless functions experience "cold starts"~\cite{vahidinia2020cold, silva2020prebaking} 
when provisioning new execution environments, 
introducing latency and additional resource consumption. 
Malicious actors could exploit this characteristic through coordinated attacks using distributed bots to simultaneously trigger numerous cold starts. 
This would degrade performance and potentially inflate costs. 
As Ahmadi et~al. note, 
mitigating such attacks requires limiting external access during cold start phases to enhance reliability~\cite{ahmadi2024challenges}. 
Our design partially addresses this attack through minimal function code size and function refreshing, 
but remains vulnerable to sophisticated distributed attacks.

%% file: sections/6_relatedworks.tex



%% file: sections/7_futurework.tex
\section{Future Work}


Future work in this area could explore several promising directions. Performance enhancements could be developed to improve \sys's responsiveness under varying network conditions, including high-latency or congested networks common in censored regions. 
Further research could also explore adaptive bridge allocation strategies that dynamically respond to changes in censorship patterns and intensity. By collecting and analyzing real-time data on connection failures and successful circumventions, \sys could intelligently distribute traffic across regions and providers to maximize availability while minimizing costs. 



Given that \sys functions on a request-response model, it presents opportunities for integration with other anti-censorship ecosystems. For example, it could be applied to Tor bridge or Shadowsocks airport~\cite{chua_fighting_2019} distribution, serving as lightweight entry points into anonymity networks or anti-censorship systems. Additionally, \sys could facilitate the secure distribution of the latest circumvention tool binaries (e.g., Tor Browser~\cite{torprojectProjectPrivacy}, Psiphon~\cite{githubGitHubPsiphonIncpsiphonautomation}, Lantern~\cite{githubLantern}). 


\section{Conclusion}
In this work, 
we present \sys, 
a novel serverless computing-based censorship circumvention system that inherits the benefits of serverless computing. 
Our approach addresses the fundamental challenges of existing circumvention proxies by prioritizing cost efficiency, censorship resistance, and performance.

\sys demonstrates that serverless functions offer significant advantages for censorship circumvention compared to traditional cloud-based approaches. Utilizing the ephemeral nature of serverless computing, 
our system can rapidly deploy and rotate bridges across different regions, 
making it difficult for censors to identify and block by IP address and domain name. 
The stateless design of serverless functions enables seamless migration between bridges without service interruption, 
providing continuous connectivity for users in censored regions. 
Additionally, we provide a privacy-preserving mode of \sys that supports tunneling over SOCKS with a cost-privacy trade-off.
Notably, \sys can achieve 97\% of cost savings through serverless computing's pay-as-you-go pricing based on the number of requests.

\section{Ethical Considerations}
For the real-world deployment evaluation of \sys in the censored region, we sent multiple requests to the known-blocked websites. In the whole experiment process, we adhere to comply with ethical standards. To minimize the risk, measurements were performed from an Alibaba Cloud server under our control rather than using personal devices or residential connections. Mitigating possible harm to the cloud provider, requests were limited to non-interactive curl probes checking response status only. Therefore, no participants were put at risk during our evaluation.

The vanilla \sys implementation reassembles HTTPS packets at serverless bridges, creating a trust relationship between users and bridge operators who have technical access to user traffic. We explicitly acknowledge this limitation and provide a privacy-preserving mode with encrypted channels at increased cost. As a consequence of \sys being available to the public, careful evaluation on operator's trustworthiness by users is required. We recommend operating serverless bridges for individual purposes or by trustworthy organizations with strict no-logging policies. 

\section{Artifact Availability}
Our \sys source code is publicly available at: 
\href{https://anonymous.4open.science/r/CensorLess-08DB/}{\path{https://anonymous.4open.science/r/CensorLess-08DB/}}

%% file: sections/appendix.tex

\section{Additional Experiments}
\subsection{\sys Performance}~\label{appendix:eval-performance}
\begin{figure*}[t]
    \centering
    \subfloat[cnn.com browsing: without proxy\label{fig:sub1}]{%
        \includegraphics[width=0.32\textwidth]{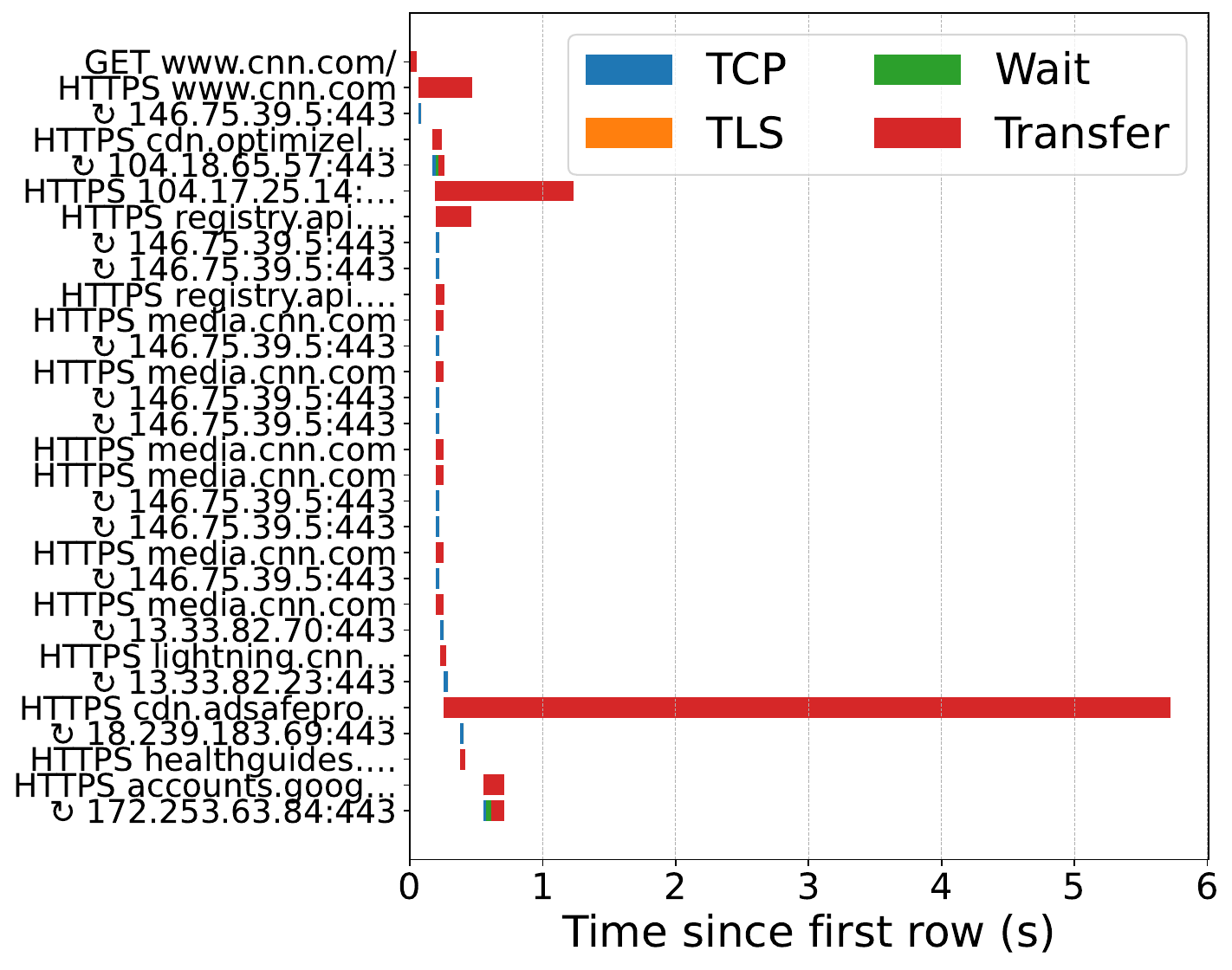}%
    }\hfill
    \subfloat[PDF downloading: without proxy\label{fig:sub2}]{%
        \includegraphics[width=0.32\textwidth]{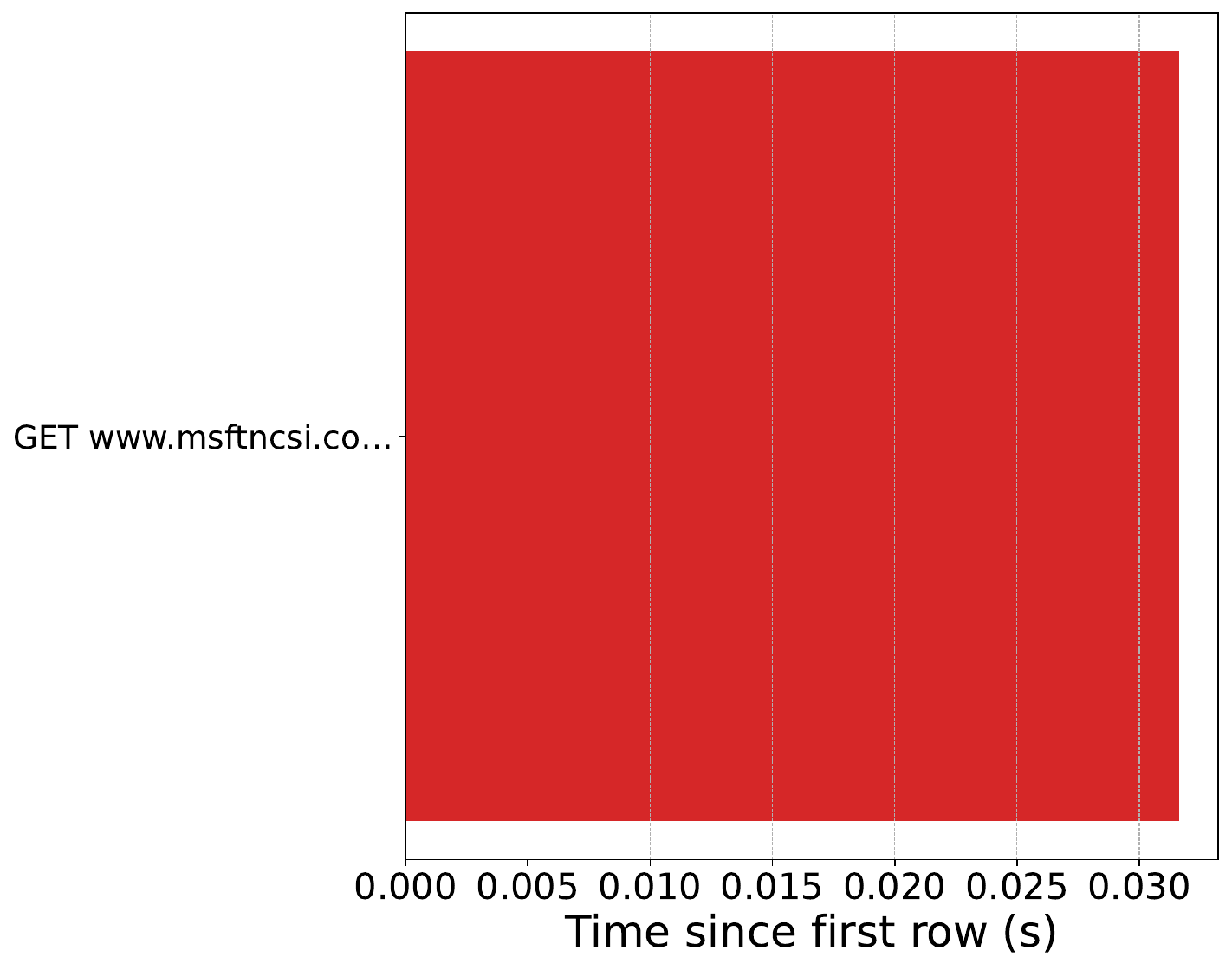}%
    }\hfill
    \subfloat[Watching Video: without proxy\label{fig:sub3}]{%
        \includegraphics[width=0.32\textwidth]{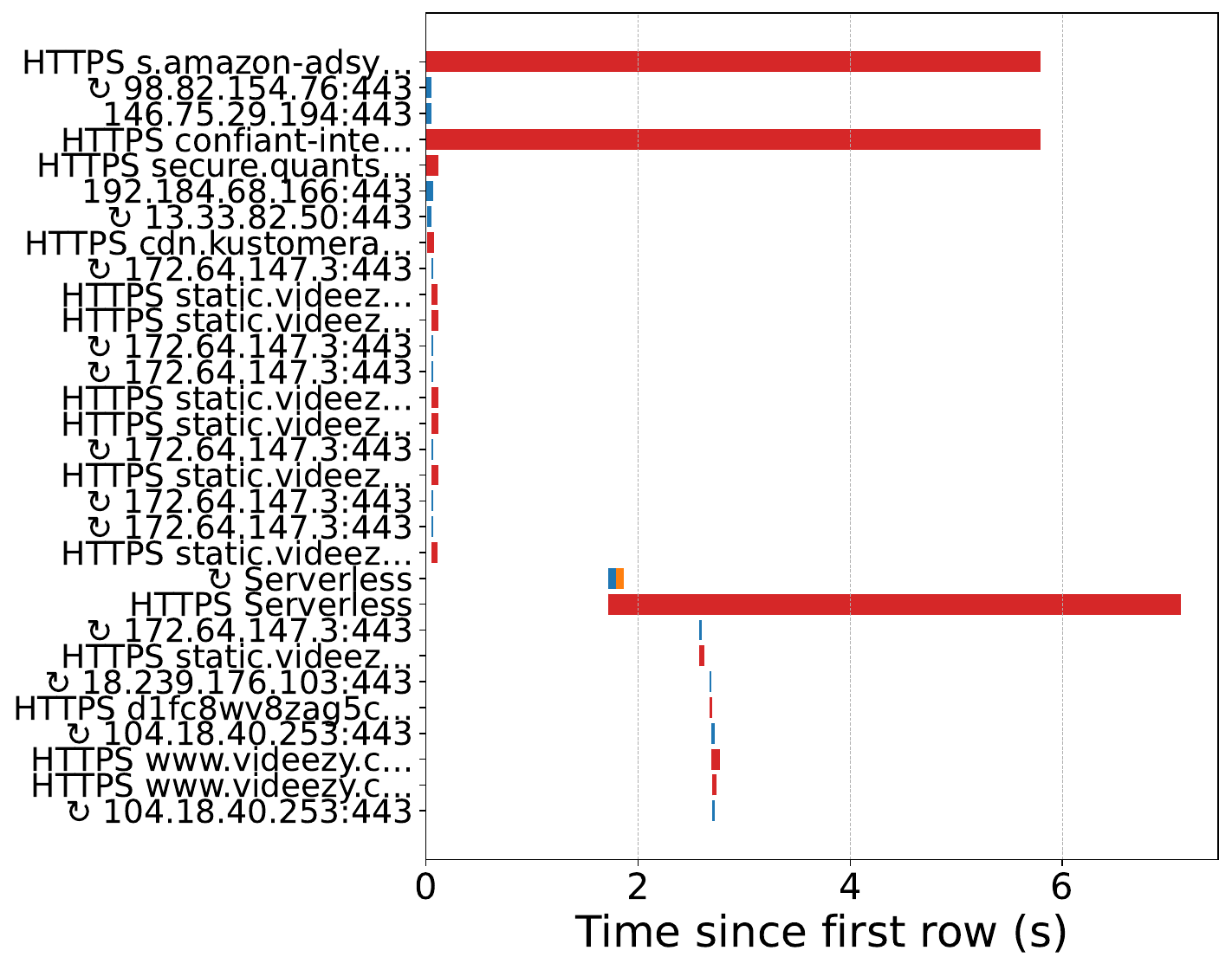}%
    }
    
    
    \subfloat[cnn.com browsing: vanilla\label{fig:sub4}]{%
        \includegraphics[width=0.32\textwidth]{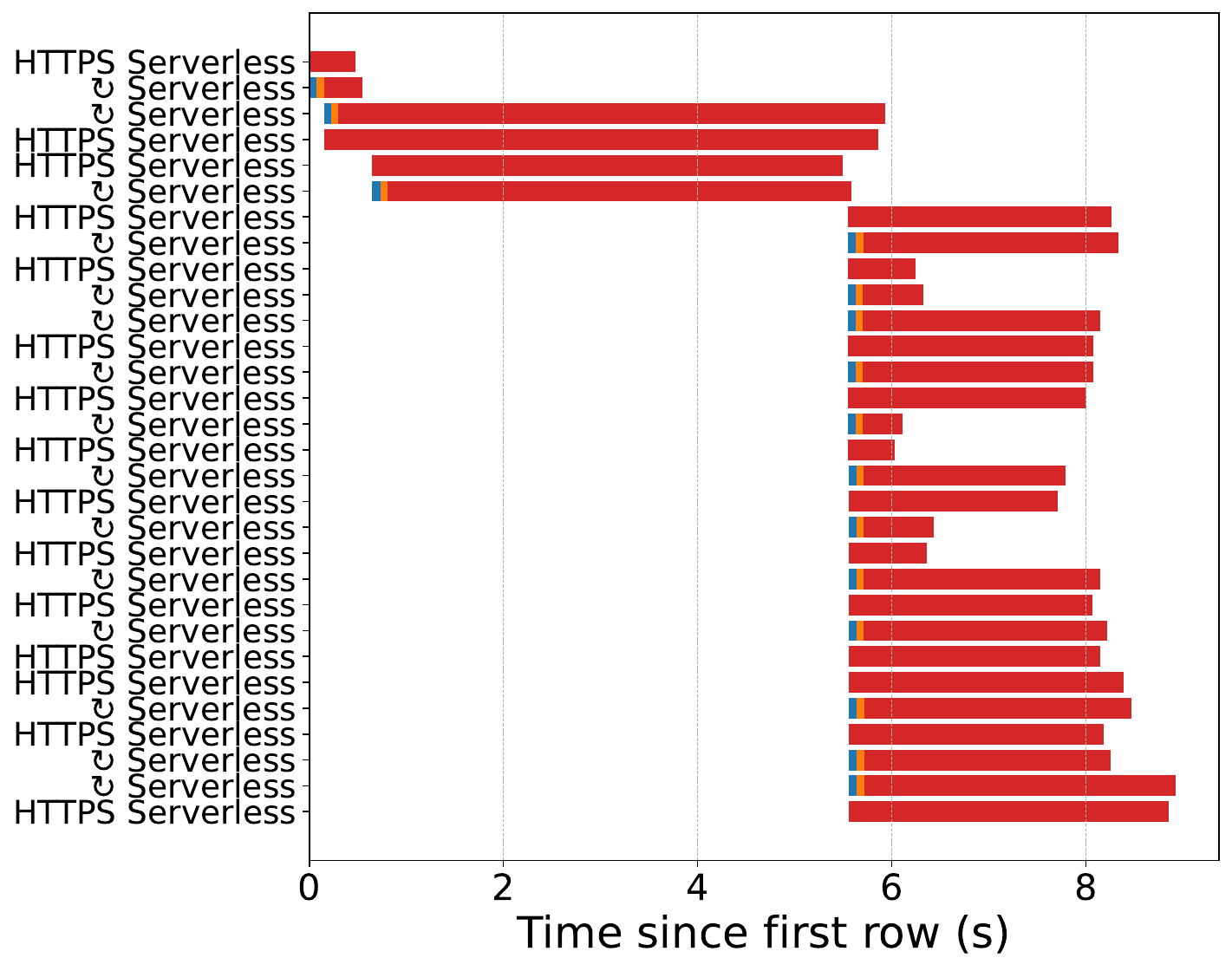}%
    }\hfill
    \subfloat[PDF downloading: vanilla\label{fig:sub5}]{%
        \includegraphics[width=0.32\textwidth]{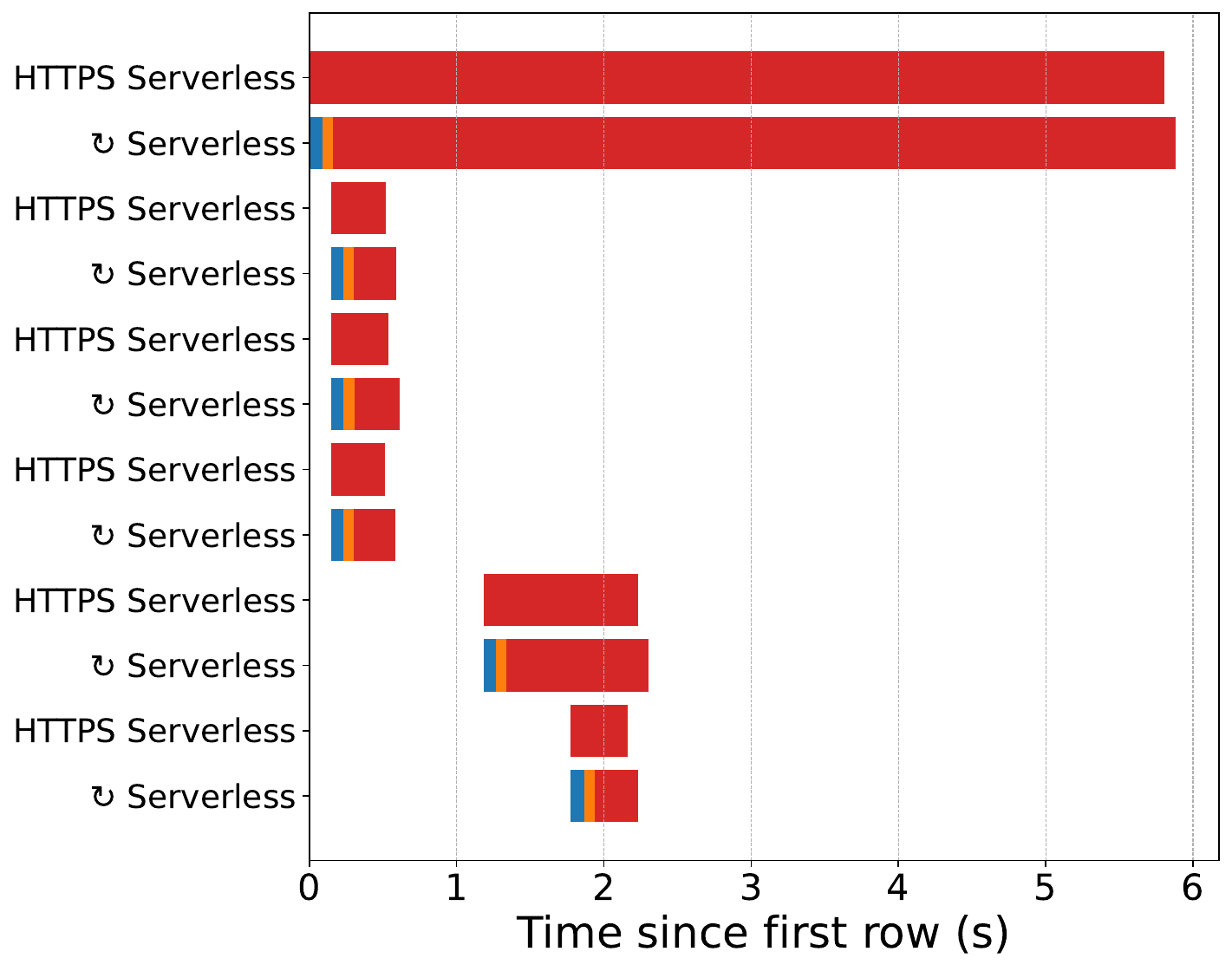}%
    }\hfill
    \subfloat[Watching Video: vanilla\label{fig:sub6}]{%
        \includegraphics[width=0.32\textwidth]{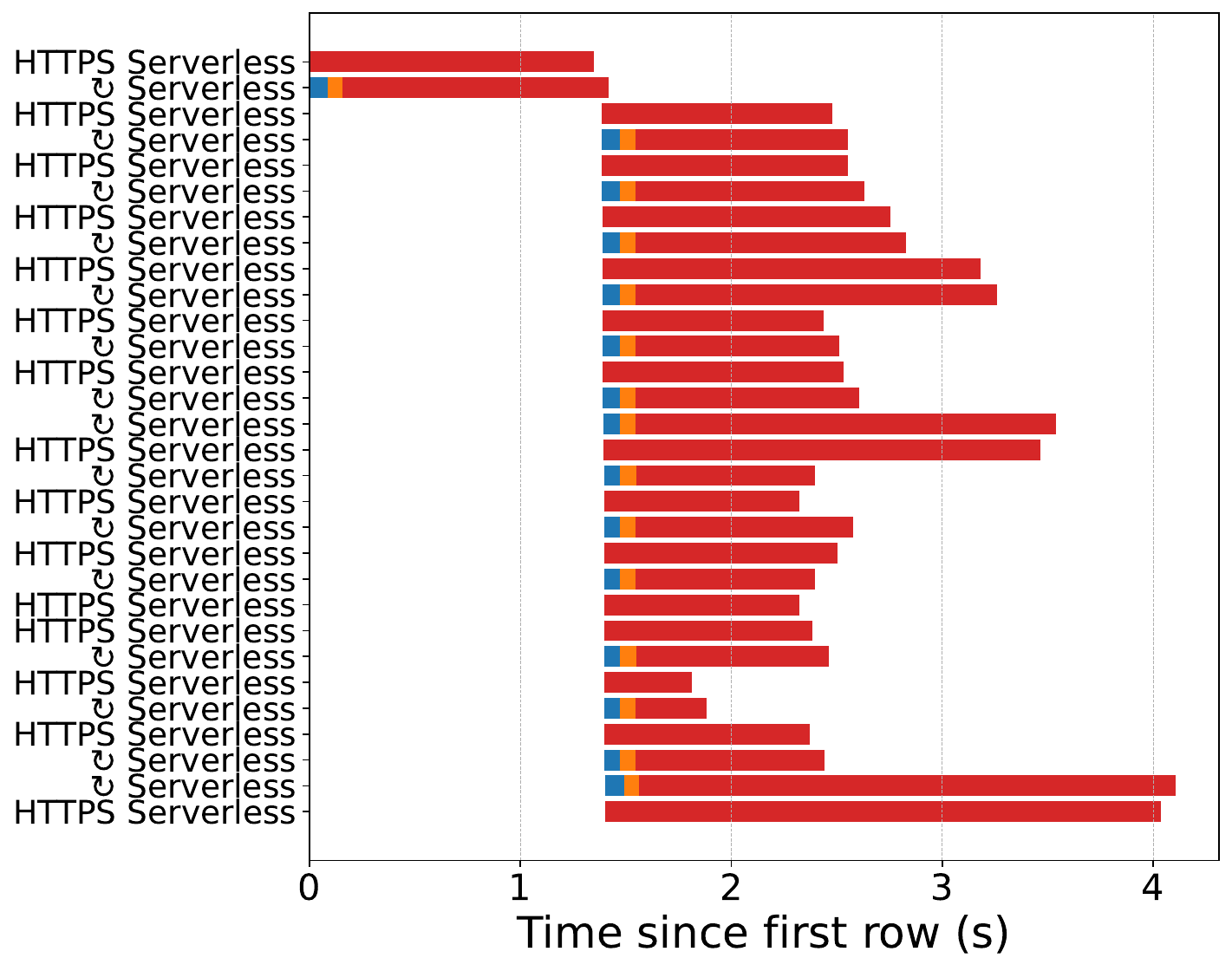}%
    }
    
    
    \subfloat[cnn.com browsing: vanilla with migration\label{fig:sub7}]{%
        \includegraphics[width=0.32\textwidth]{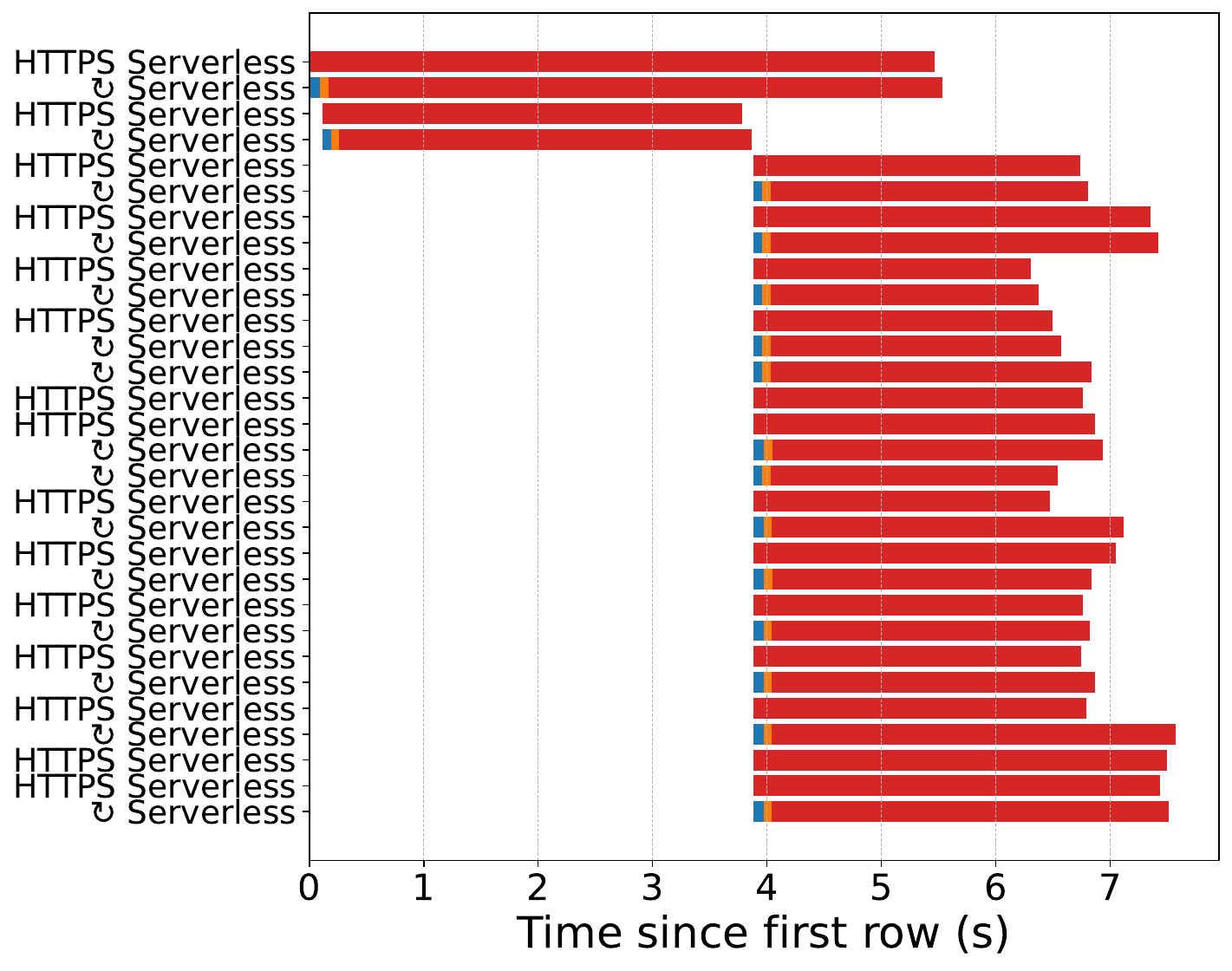}%
    }\hfill
    \subfloat[PDF downloading: vanilla with migration\label{fig:sub8}]{%
        \includegraphics[width=0.32\textwidth]{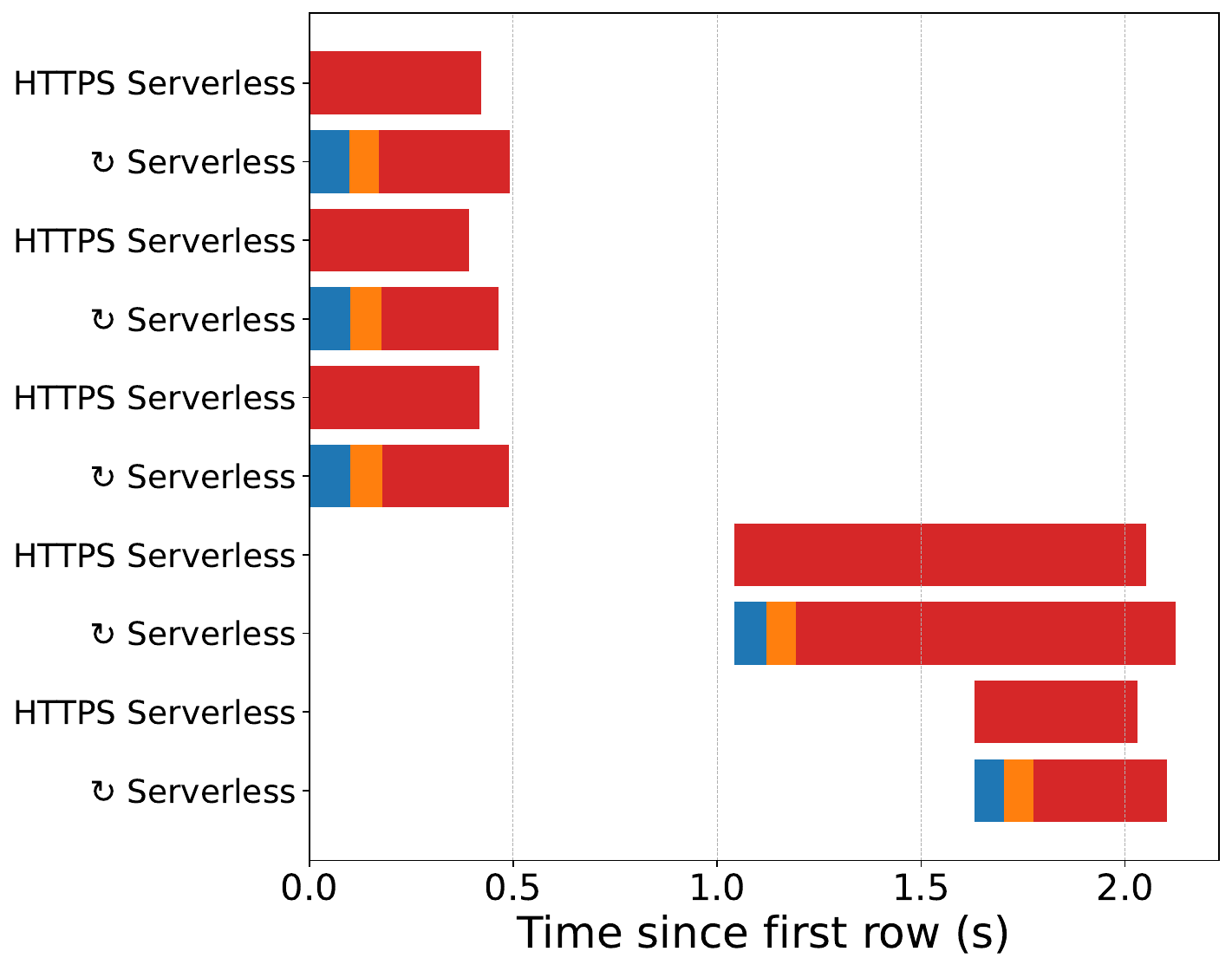}%
    }\hfill
    \subfloat[Watching Video: vanilla with migration\label{fig:sub9}]{%
        \includegraphics[width=0.32\textwidth]{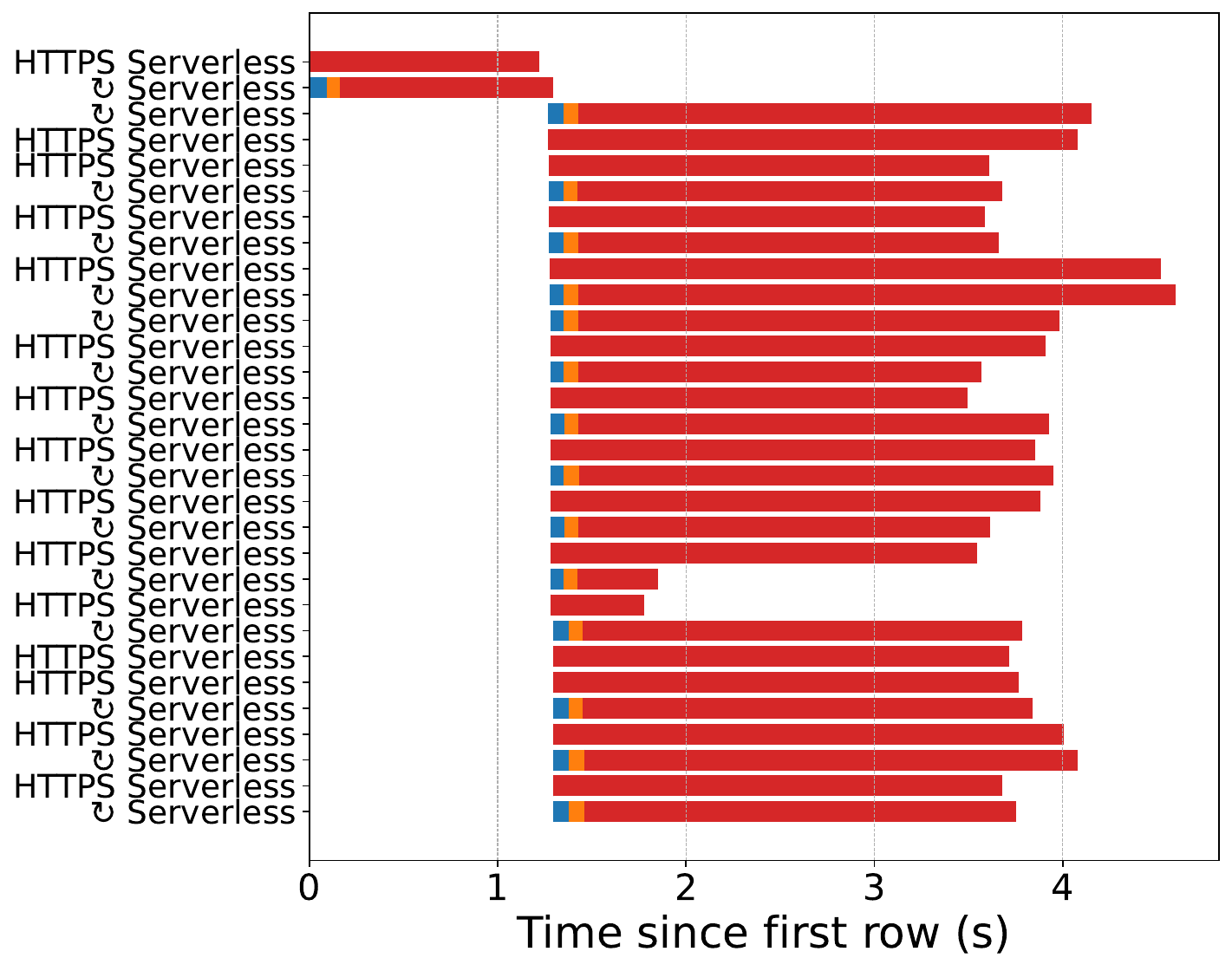}%
    }
    
    
    \subfloat[cnn.com browsing: private mode\label{fig:sub10}]{%
        \includegraphics[width=0.32\textwidth]{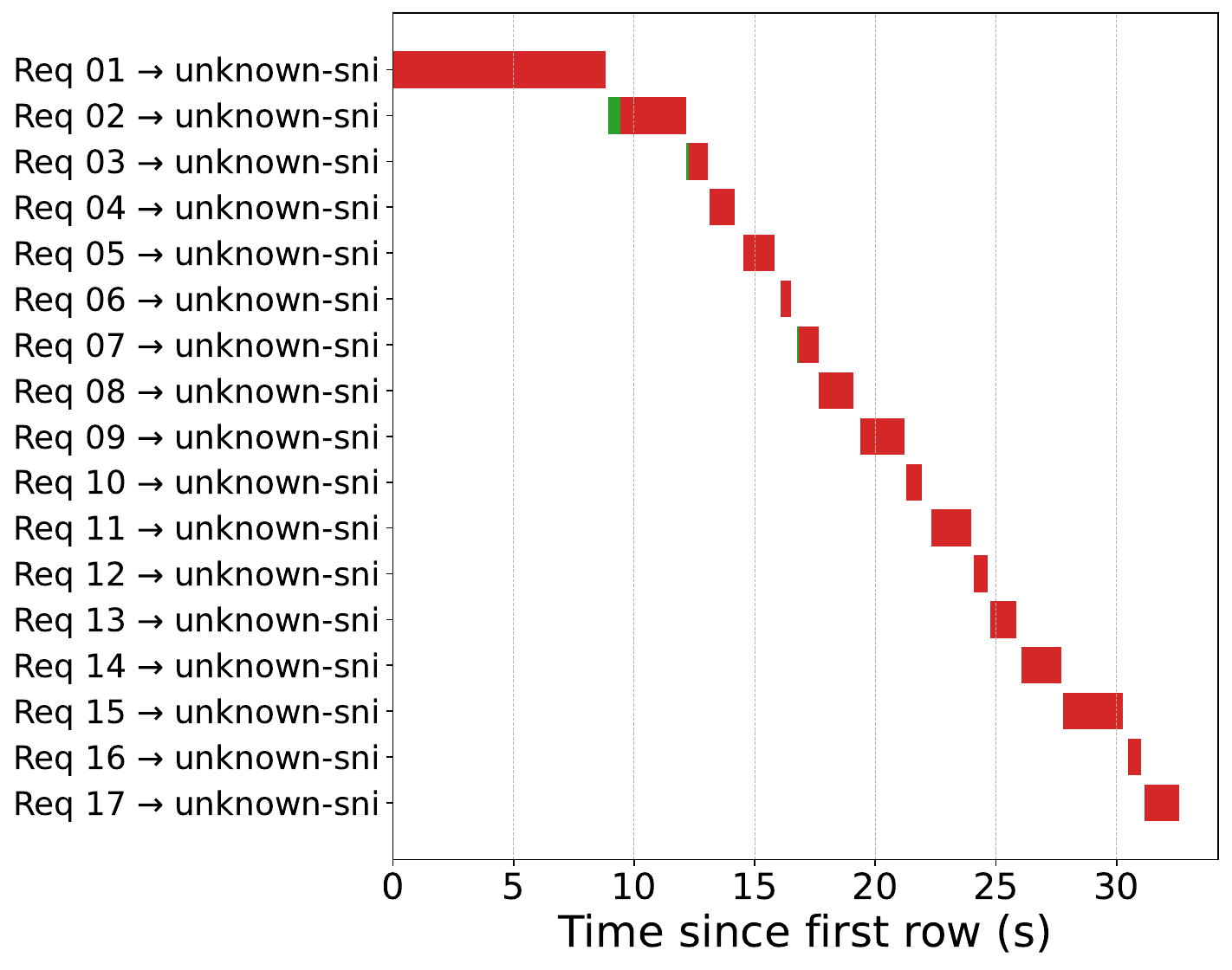}%
    }\hfill
    \subfloat[PDF downloading: private mode\label{fig:sub11}]{%
        \includegraphics[width=0.32\textwidth]{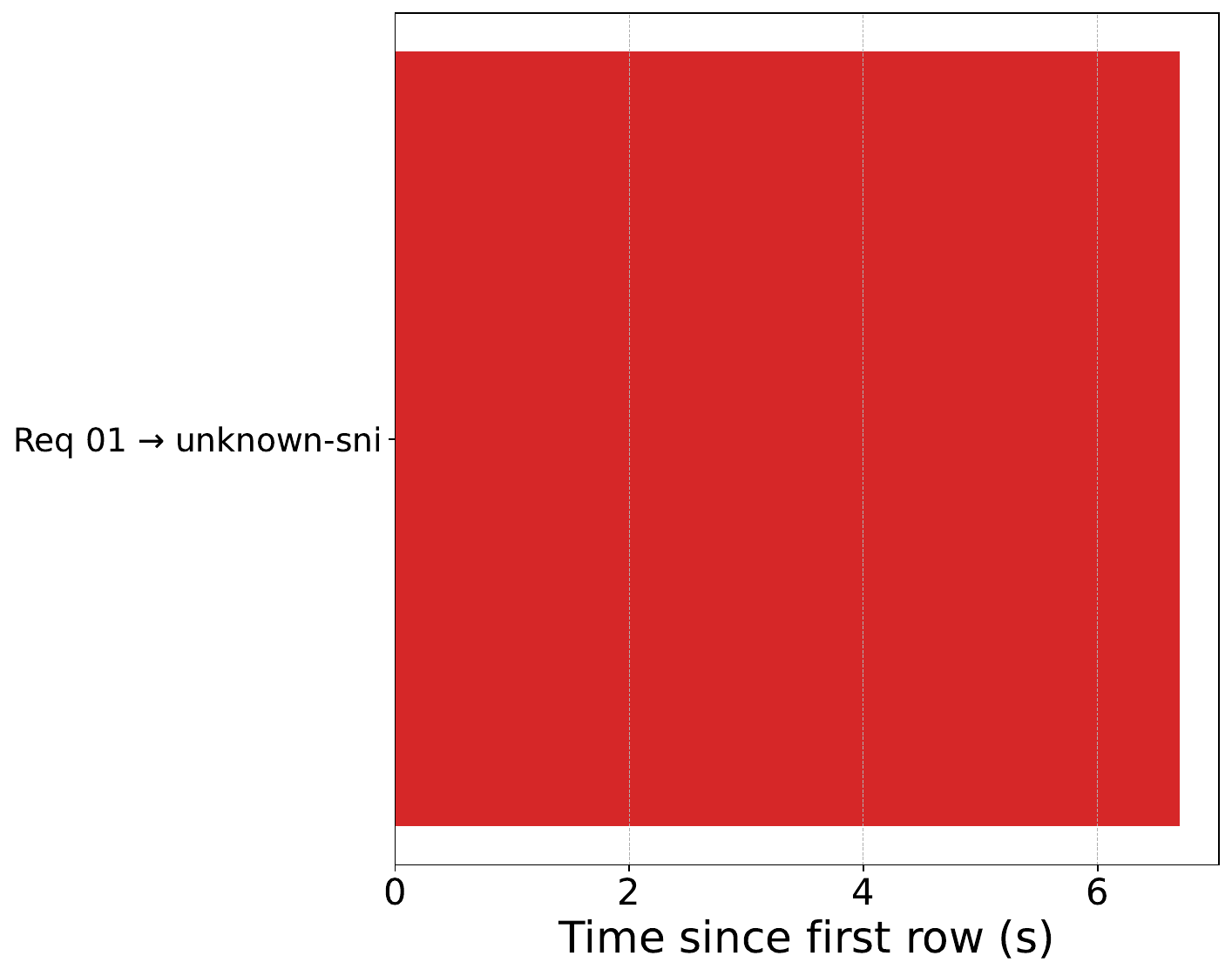}%
    }\hfill
    \subfloat[Watching Video: private mode\label{fig:sub12}]{%
        \includegraphics[width=0.32\textwidth]{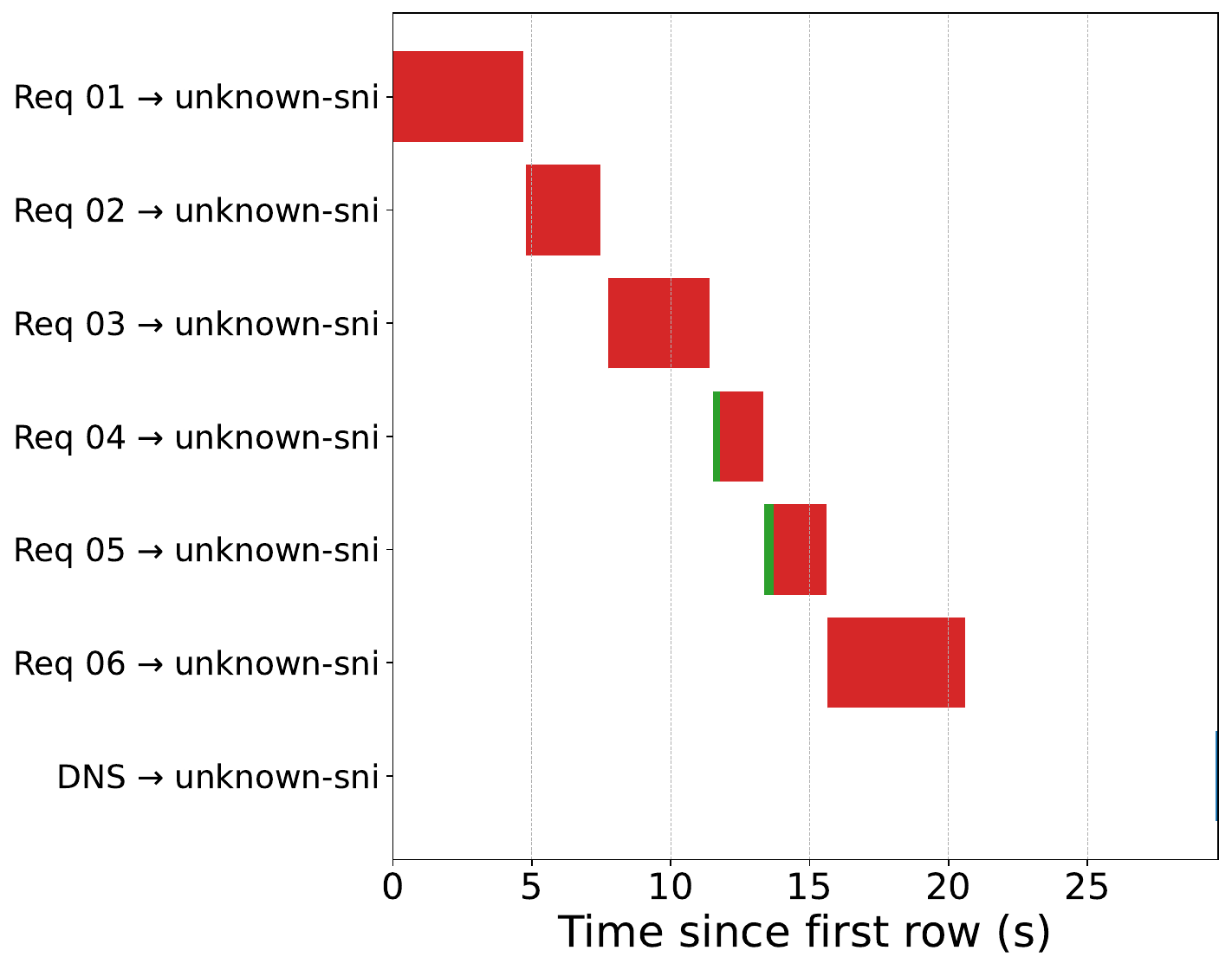}%
    }
    
    \caption{Baseline comparison for three use cases: loading the cnn.com page, which has many objects, downloading a PDF file, and watching a short video.}
    \label{fig:waterfall}
\end{figure*}

In Figure~\ref{fig:waterfall}, each subfigure shows the time to load the content in three different cases: browsing the cnn.com webpage, downloading a PDF file, and watching a short video. The Y label exhibits the destination where the request goes, and $\circlearrowright$ indicates the reuse connection to the same destination. While the waterfall result without a proxy can specify the destination, the result with \sys successfully hides the destination of requests. Vanilla mode \sys without and with bridge migration have similar load time patterns because they use HTTPS GET/POST methods only. In Figure~\ref{fig:sub8}, vanilla with migration seamlessly handles the client requests even though the bridge migration happened between 0.5 and 1.0 second. Notably, \sys private mode requires more time to load the webpage because it routes traffic to an additional VPS and goes through a connection setup process for secure communication.

\subsection{Operational Costs}~\label{appendix:eval-cost}
\begin{figure}[h]
  \centering
  \includegraphics[width=\linewidth]{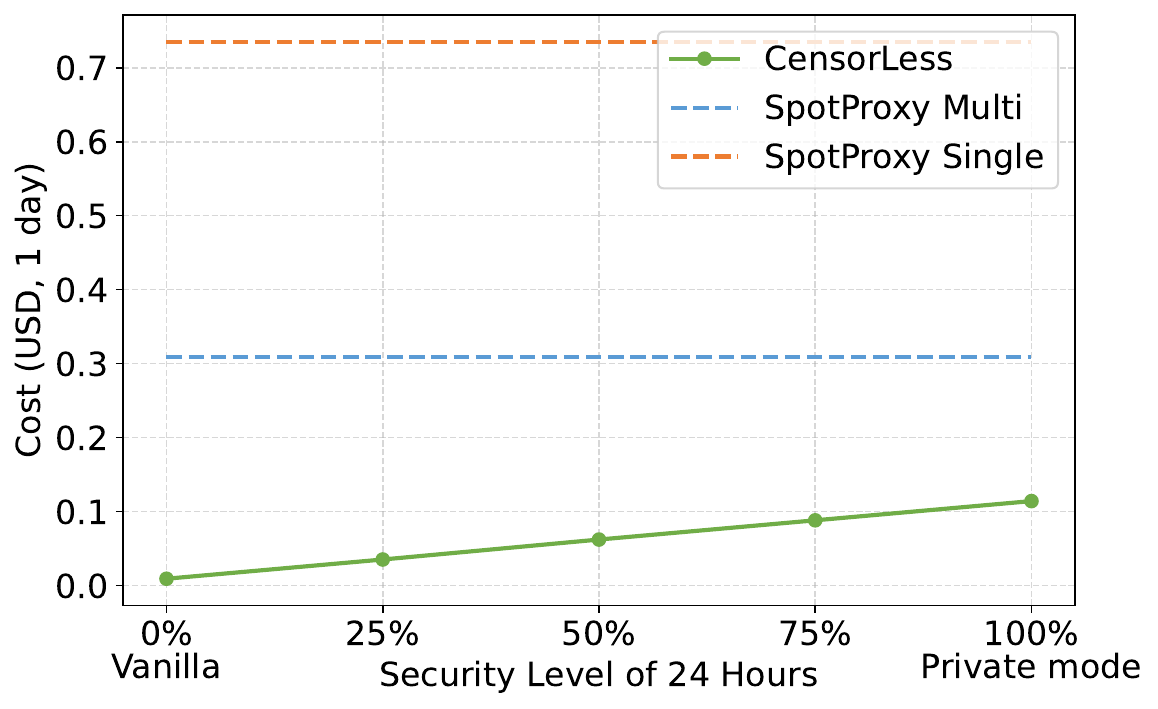}
  \caption{\textbf{Cost comparison result by the security level of a single proxy.} 25\% means that 25\% of the total traffic over the 24 hours (6 hours) is used in private mode.}
  \label{fig:cost-sec-lev}
\end{figure}

We analyzed how the operational cost of a single proxy changes depending on the security level. Since the private mode incurs an additional cost for the VPS based on hourly usage, we measured the security level as the number of hours of using the private mode. 0\% of security level indicates the client uses only vanilla \sys for 24 hours, 25\% indicates the client uses the private mode for 6 hours out of 24 hours, and 100\% indicates the entire traffic for 24 hours is in private mode. 
While the state-of-the-art cost-efficient tool, SpotProxy, uses spot instances (cost-saving IaaS instances) for individual proxies, vanilla \sys uses request-based serverless functions for individual proxies and additionally adopts the IaaS instance (i.e., VPS) for the private mode. In our experiment, this VPS requires a lower configuration (2~CPUs and 1~GB memory) compared to SpotProxy (2~CPUs and 8~GB memory), resulting in a lower hourly cost. Figure~\ref{fig:cost-sec-lev} demonstrates a gradual and small linear increase in costs as the usage of the private mode increases.  Even at 100\% (i.e., private mode for 24 hours), the security level charges less than SpotProxy's cheapest scenario. The cost of the private mode primarily depends on the VPS cost, similar to SpotProxy; however, a single VPS in private mode supports multiple serverless proxies, saving costs as the number of proxies increases.

When analyzing serverless function economics at scale, we discovered that cost linearity breaks at higher concurrency levels due to varying function durations.
Table \ref{tab:client-duration} shows that lower invocation numbers actually consume more time due to cold starts, affecting overall costs since duration impacts billing.


\begin{table}[!t]
    \caption{Average Serverless Function Duration by Different Number of Concurrent Invocations}
    \label{tab:client-duration}
    \centering
    \renewcommand{\arraystretch}{1.3}
    \begin{tabular}{lc}
        \toprule
        \textbf{Invocations} & \textbf{Duration (ms)} \\
        \midrule
        10  & 4557.69 \\
        50  & 1002.35 \\
        100 & 1863.76 \\
        200 & 720.85 \\
        \bottomrule
    \end{tabular}
\end{table}

\begin{figure}[h]
  \centering
  \includegraphics[width=\linewidth]{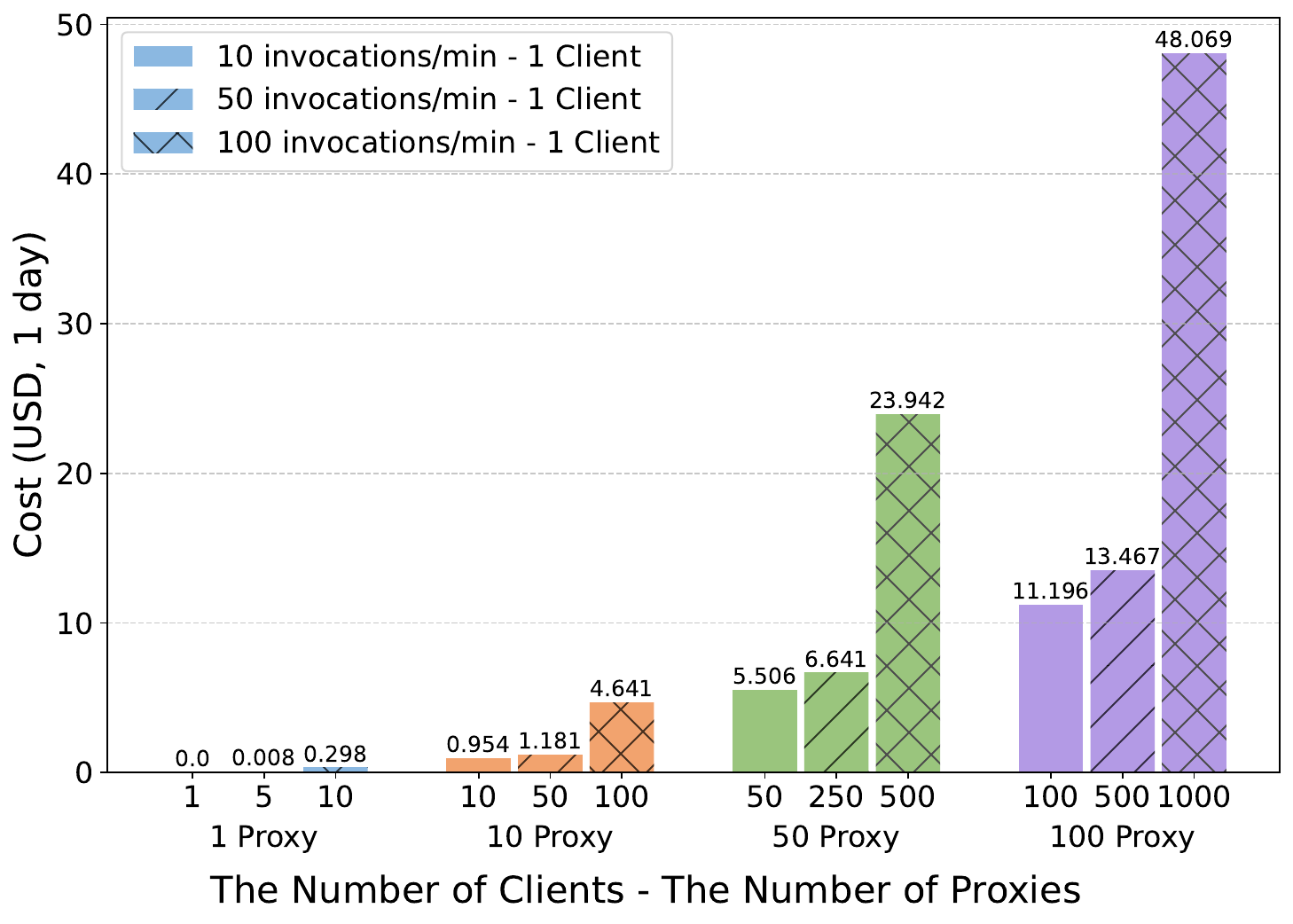}
  \caption{\textbf{Serverless bridge cost scaling (1~day) results when the number of clients and proxies differ.} Bars in each group indicate the different numbers of clients. Costs are compared when the number of invocations sent per minute by one client within each group varies to 10, 50, and 100.}
  \label{fig:cost-per-proxy}
\end{figure}

We investigated the serverless bridge operational cost as the number of proxies and clients increases. 
In Figure~\ref{fig:cost-per-proxy}, 
each bar in the same position within each group represents the same number of invocations when a client sends to one proxy. 
Due to the relatively long duration at the lower invocations, 
the costs of 10 invocations/client and 50 invocations/client are similar. The serverless function charges a price increase of around 3.6 times when a client sends 100 requests to one proxy. 
This means maintaining moderate invocation levels (around 50 per client) with more proxies is more cost-efficient than fewer proxies handling burst traffic, when the total number of function calls remains equal. 
This insight provides valuable guidance for optimal system deployment in practice.

\subsection{Circumvention Effectiveness}~\label{appendix:experiment}
\subsubsection{\textbf{Simulation.}}~\label{appendix:eval-simulation}
\begin{figure}[h]
  \centering
  \includegraphics[width=\linewidth]{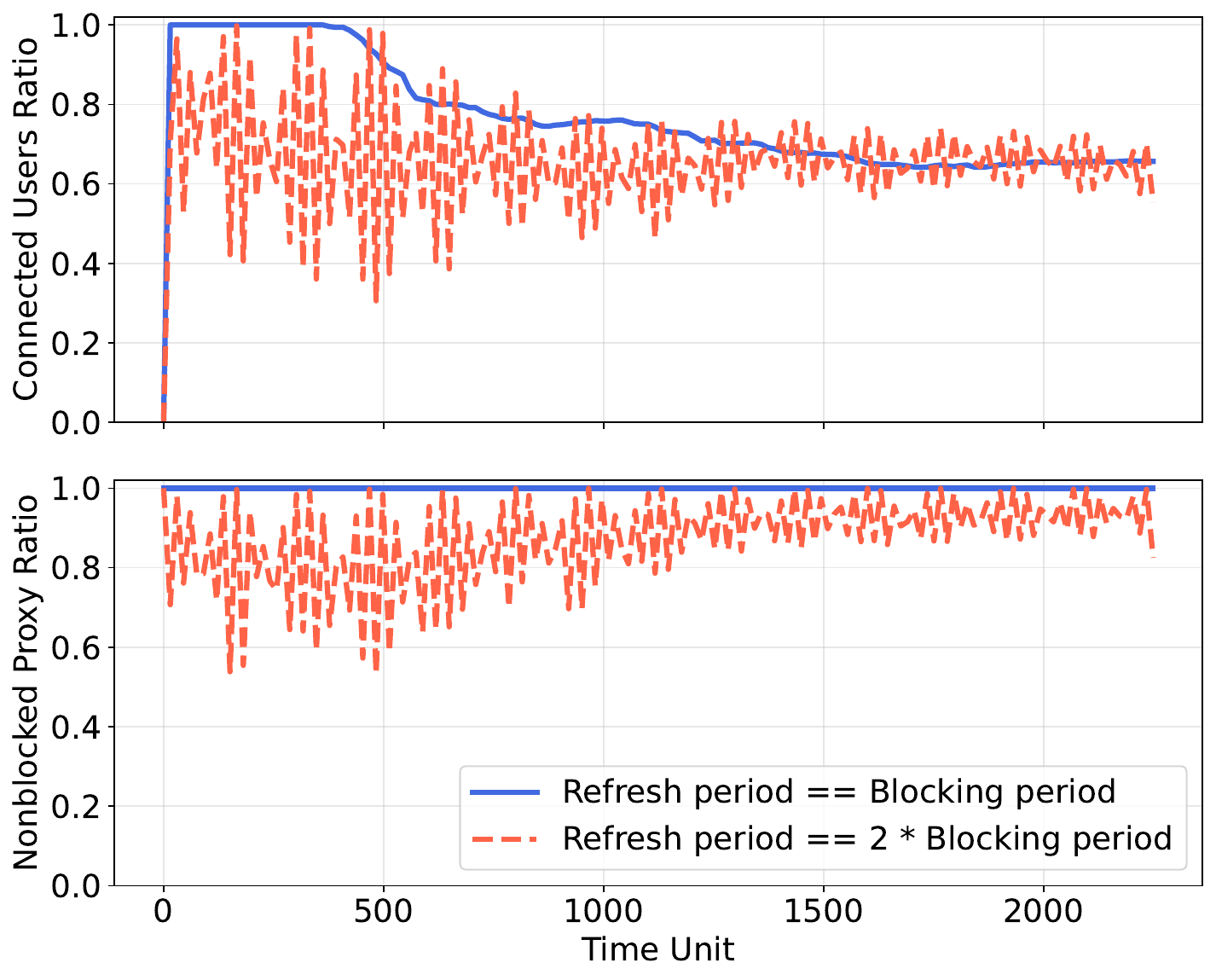}
  \caption{\textbf{Censor simulation using the Aggressive method.} \sys preserves 66\% of connected users and 87\% of nonblocked proxies, even when 50\% of total clients are censor agents (when the refresh cycle is twice the censorship cycle).}
  \label{fig:simulation-aggressive}
\end{figure}

\begin{figure}[h]
  \centering
  \includegraphics[width=\linewidth]{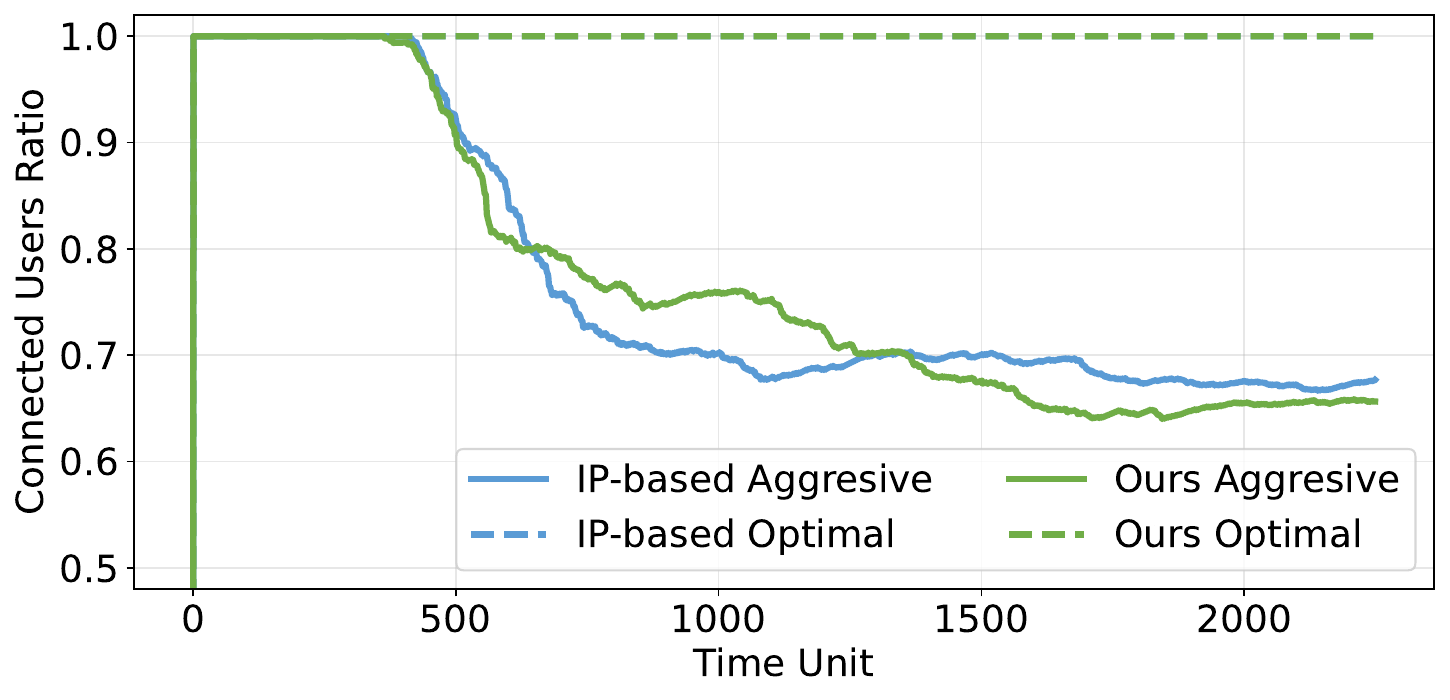}
  \caption{\textbf{Censor simulation comparison.} URL-based (i.e., \sys) and IP-based proxy rotating tool present similar censor resistance results (77\% connected users with the Aggresive method) when 50\% of total clients are censor agents and the refresh cycle is the same as censor cycle.}
  \label{fig:simulation-compare}
\end{figure}

We evaluated our system's circumvention effectiveness on both the client and proxy sides against two types of censors: aggressive and optimal. Aggressive censor agents immediately block any new proxy that they identify, while the optimal censor uses an optimal game-theoretic approach as its censorship strategy. Setting 50\% of clients as censor agents, we compared scenarios where refresh periods were either equal to or double the blocking periods. As shown in Figure~\ref{fig:simulation-aggressive}, our system shows large fluctuations in the initial phases for both connected users and nonblocked proxy ratios when the refresh period was set to double the blocking period, while matching the refresh period with the blocking period shows stable connections. Even with 50\% of users being aggressive censor agents, \sys guarantees an average of 70\% connectivity to users.

\sys demonstrates censorship circumvention efficacy comparable to the state-of-the-art IP-based tool (e.g., SpotProxy), as it resists censorship by rotating proxies using URLs. In this simulation, the censor blocks users and proxies with identified URLs, following the same approach as the IP-based proxy assignment algorithm. As shown in Figure~\ref{fig:simulation-compare}, although the dominating algorithm varies over time, both provide identical ratios of connected users on average---77\% with the aggressive method and 99\% with the optimal method. 

\subsubsection{\textbf{In Censored Region.}}
\begin{table}[!t]
    \caption{Complete website access results of \sys in the censored region, Nanjing, China}
    \centering
    \renewcommand{\arraystretch}{1.3}
    \label{tab:appendix-exp}
    \resizebox{\columnwidth}{!}{
    \begin{tabular}{lcc|lcc}
    \toprule
    \textbf{Websites}& \textbf{Vanilla} & \textbf{Private} & \textbf{Websites}& \textbf{Vanilla} & \textbf{Private} \\
    \midrule
    jpc.de & \checkmark & \checkmark & huffpost.com & \checkmark & \checkmark\\
    ddd-smart.net & \checkmark & \checkmark & pornstars.tube & \checkmark & \checkmark\\
    google.com.et & \checkmark & \checkmark & fosstodon.org & \checkmark & \checkmark\\
    njav.tv & \checkmark & \checkmark & lohaco.jp & \checkmark & \checkmark\\
    popyard.space & \checkmark & \checkmark & asiafinancial.com & \checkmark & \checkmark\\
    google.co.hu & \checkmark & \checkmark & red-movies.com & \checkmark & \checkmark\\
    chatdoc.com & \checkmark & \checkmark & nifty.org & \checkmark & \checkmark\\
    dergipark.org.tr & \checkmark & \checkmark & domai.com & \checkmark & \checkmark\\
    ero-labs.com & \checkmark & \checkmark & factmandu.com & \checkmark & \checkmark\\
    wikipedia.org & \checkmark & \checkmark & rfa.org & \checkmark & \checkmark\\
    nationalfile.com & \checkmark & \checkmark & getlink.pro & \checkmark & \checkmark\\
    gofundme.com & \checkmark & \checkmark & erotic-hentai.com & \checkmark & \checkmark\\
    indiandefencereview.com & \checkmark & \checkmark & thecitizen.in & \checkmark & \checkmark\\
    thumbnailseries.com & \checkmark & \checkmark & simplex.im & \checkmark & \checkmark\\
    voacantonese.com & \checkmark & \checkmark & tubegays.xxx & \checkmark & $\times$\tablefootnote{It indicates the system did not receive any response from the website.}\\
    hostux.net & \checkmark & \checkmark & beliefnet.com & \checkmark & \checkmark\\
    flyingjizz.com & \checkmark & \checkmark & mastodon.world & \checkmark & \checkmark\\
    gamestorrents.fm & \checkmark & \checkmark & xfrenchies.com & \checkmark & \checkmark\\
    epochtimes.com.ua & \checkmark & \checkmark & aflegal.org & \checkmark & \checkmark\\
    ejecentral.com.mx & \checkmark & \checkmark & aiweiwei.com & \checkmark & \checkmark\\
    flyflv.com & \checkmark & \checkmark & voyeurweb.com & \checkmark & \checkmark\\
    voachinese.com & \checkmark & \checkmark & memeorandum.com & \checkmark & \checkmark\\
    swag.live & \checkmark & \checkmark & chat-gpt.org & \checkmark & \checkmark\\
    modelhub.com & \checkmark & \checkmark & newsdirectory3.com & \checkmark & \checkmark\\
    ideapocket.com & \checkmark & \checkmark & hsav.xyz & \checkmark & \checkmark\\
    \bottomrule
    \end{tabular}
    }
    \label{appendix:tab:china-censor}
\end{table}

We experimented with 500 blocked websites in mainland China, a well-known censored region.
The part of the results, against 50 blocked domains, is shown in Table~\ref{appendix:tab:china-censor}. The vanilla \sys successfully receives a response for every request sent to those domains, while the private mode only fails to receive responses from tubegays.xxx. The failure of tubegays.xxx in private mode is due to AWS VPS's request being denied by the website policy. While the final request to the website is from AWS VPS, the vanilla mode \sys successfully received the response since it uses serverless function. 